\documentclass[a4paper,11pt]{article}
\pdfoutput=1 

\usepackage{jheppub}  
\usepackage{graphicx,color}
\usepackage{amsmath}
\usepackage{autobreak}
\allowdisplaybreaks
\newcommand{\ben}{\begin{enumerate}}
\newcommand{\een}{\end{enumerate}}
\newcommand{\beq}{\begin{equation}}
\newcommand{\eeq}{\end{equation}}
\newcommand{\dFA}{\frac{d_{FA}^{(4)}}{N_F}}
\newcommand{\dFF}{\frac{d_{FF}^{(4)}}{N_F}}
\def\g0#1bbH{{g^{b(#1)}_{_{d,0}} }}
\def\gSVN#1{\Delta^{(#1)}_{d,b\bar{b}}}
\usepackage[dvipsnames]{xcolor}
\newcommand{\LNoNtb}{{\color{BlueViolet}\overbar{\boldsymbol{L}}}}

\def\z#1{\zeta_{#1}}
\newcommand{\CA}{C_A}
\newcommand{\CF}{C_F}
\newcommand{\NF}{n_f}

\newcommand{\Lqr}{L_{qr}}
\newcommand{\Lfr}{L_{fr}}
\newcommand{\Lqrfr}{L_{qr}L_{fr}}

\def\bt#1{{\beta_{#1}}}

\def\z#1{\zeta_{#1}}
\def\nf{n_f}

\def\Cf{C_{F}}

\def\Ca{C_{A}}

\def\dFAbN{{\frac{d_{FA}}{N_c}}}

\newcommand{\eq}[1]{eq.\ (\ref{#1})}
\newcommand{\fig}[1]{fig.\ (\ref{#1})}

\newcommand{\sect}[1]{sec.\ (\ref{#1})}
\newcommand{\app}[1]{\ (\ref{#1})}

\definecolor{amber}{rgb}{1.0, 0.49, 0.0}
\newcommand{\uncert}[2][]{{$^{#1}_{#2} \%$}}
\newcommand{\nn}{\nonumber\\}
\newcommand{\df}{{\rm{d}}}
\newcommand{\mur}{\mu_r}
\newcommand{\muf}{\mu_f}
\def\w{\overbar{\omega}}
\newcommand{\overbar}[1]{\,\overline{\!{#1}}}
\newcommand{\Nbar}{\overbar{N}}
\newcommand{\Nbari}{\overbar{N}_i}
\def\Nbar#1{\overbar{N}_{#1}}
\newcommand{\as}{a_S}
\newcommand{\als}{\alpha_S}
\newcommand{\lam}{\lambda}
\def\gm#1{{\gamma^{(#1)}_m}}

\title{\bf Higgs boson rapidity distribution in bottom annihilation
	at NNLL and beyond}
\author[]{Goutam Das}
\affiliation[]{
	Institut f{\"u}r Theoretische 
	Teilchenphysik und Kosmologie,\\
	RWTH Aachen University, 
	D-52056 Aachen, Germany
}
\emailAdd{goutam@physik.rwth-aachen.de}

\abstract{
We present precise resummed predictions for Higgs boson 
rapidity distribution through bottom quark annihilation 
at next-to-next-to-leading logarithmic (NNLL) accuracy 
matched to next-to-next-to-leading order (NNLO) and at
next-to-next-to-next-to-leading logarithmic (N3LL) accuracy 
matched 
to next-to-next-to-next-to-leading order soft-virtual 
(N3LOsv) in the strong coupling. Exploiting the universal behavior of 
soft radiation near the threshold, we determine the 
analytic expressions for the process-dependent and 
universal perturbative ingredients for threshold 
resummation in double singular limits of partonic 
threshold variables $z_1,z_2$. Subsequently, the 
threshold resummation is performed in the double 
Mellin space within the standard QCD framework. 
The new third-order process-dependent 
non-logarithmic coefficients are determined
using three-loop bottom 
quark form factor and third-order quark soft distribution 
function in rapidity distribution. 
The effect of these new resummed coefficients are 
studied at the $13$ TeV LHC.
We observe a better perturbative convergence in the 
resummed predictions on the Higgs rapidity spectrum in 
bottom quark annihilation. We also find that the NNLL 
and N3LL corrections are sizeable which typically are 
of the order of $-2.5\%$ and $-1.5\%$ over the respective 
available fixed orders with the scale uncertainty 
remaining at the same level as the fixed order.
}

\begin{document} 
	\preprint{
		TTK-23-14, P3H-23-039
	}
	
	\maketitle
\section{Introduction} \label{sec:INTRODUCTION}
The Standard Model (SM) Higgs boson is one of the 
important fundamental particles to study at colliders 
like the Large Hadron Collider (LHC). 
Understanding the Higgs properties is crucial to 
know the SM well and is also critical to the search 
of new physics beyond the SM (BSM) where new physics 
might couple to the Higgs sector. 
Testing 
the Higgs properties and understanding its 
interactions with other fundamental particles are 
indeed main tasks at the LHC in the 
upcoming runs. 
Precision calculations play a prominent role in these 
studies by calculating the higher-order contributions 
in the perturbation theory and improving the 
predictions of Higgs boson properties to a very high 
accuracy. 

The dominant mode of Higgs production at the LHC 
is through gluon fusion. On the other hand, the 
Higgs production in bottom annihilation channel, 
despite being subdominant is also interesting to 
study. Firstly, Higgs dominantly decays to bottom 
quarks which can give direct access to the Higgs 
Yukawa coupling. This purely hadronic final state, 
however, is challenging to measure 
\cite{CMS:2018nsn,
ATLAS:2018kot}. 
This also requires 
production of Higgs boson at the first place, 
through bottom annihilation channel along with the 
dominant gluon fusion.
Secondly, it also gives access to the Higgs 
Yukawa coupling to bottom quarks even when Higgs is 
decayed through cleaner channels like di-photons 
\cite{ATLAS:2018hxb,
CMS:2020xrn} 
or four-leptons 
\cite{ATLAS:2019qet,
CMS:2021ugl} productions. 
Although in the SM the Higgs Yukawa coupling to bottom 
quark is suppressed by small bottom mass, in the 
extensions of SM \text{e.g.} two-Higgs doublet models,
or minimal supersymmetric standard model 
the coupling could be enhanced and is crucial to 
the search for new physics. Therefore, a precise 
understanding of the SM contribution will be beneficial 
in BSM analyses.
Thirdly, the Higgs production through bottom 
annihilation is also interesting on how bottom 
quark is treated: as whether it is a part of the 
proton, taking bottom quark as a massless parton 
except in the Yukawa coupling which is done in the 
5 flavor scheme (5FS), or whether it is taken as a 
massive quark throughout and excluded from the 
proton structure as is done in 4 flavor scheme (4FS). 
While in 4FS a massive bottom quark is produced from 
gluon splitting from proton, in the 5FS scheme, 
massless bottom has its own parton distributions. 
%

Due to its high importance, Higgs inclusive 
cross-section is now available theoretically to a 
very high accuracy to next-to-next-to-next-to 
leading order (N3LO) 
\cite{Anastasiou:2015vya,
Mistlberger:2018etf} 
in Higgs effective field theory (HEFT) in the gluon 
fusion channel providing a correction of around 
2\% with scale uncertainty around 3\% reducing 
from 9\% at NNLO 
\cite{Harlander:2002wh,
Anastasiou:2002yz,
Ravindran:2003um}. 
It is also known up to N3LO in the vector boson 
fusion 
\cite{Bolzoni:2010xr,
Bolzoni:2011cu,
Dreyer:2016oyx,
Buckley:2021gfw}  
where the correction already stabilizes, and the scale 
uncertainty is found to be below 0.2\% at N3LO. 
Beyond the fixed order, efforts were made to study 
the dominant threshold contributions by studying 
the soft-virtual (SV) corrections at fourth order as well 
as partial subleading logarithmic effects in gluon 
fusion \cite{Das:2019btv,Das:2020adl} with a further enhancement 
of the cross-section ranging from 
0.2\%-2.7\% depending on the scale choices. 
The fixed order cross-section is further improved by 
performing threshold resummation 
at the next-to-next-to-leading logarithmic (NNLL) 
accuracy 
\cite{Catani:2003zt,
Moch:2005ky,
Laenen:2005uz,
Idilbi:2005ni} 
and to the third logarithmic accuracy (N3LL) 
\cite{Bonvini:2014joa,
Bonvini:2016frm} 
as well as
resummation of $\pi^2$ terms \cite{Ebert:2017uel} 
arising from time-like Sudakov form factor.
The finite top mass effect is also known 
at NLO
\cite{Graudenz:1992pv,
Spira:1995rr}, 
partially at NNLO with top mass expansion 
\cite{Marzani:2008az,
Pak:2009dg,
Harlander:2009my,
Harlander:2009mq,
Pak:2009dg} 
and recently to exact NNLO \cite{Czakon:2021yub}
where an increment of 0.6\% is observed compared to 
the HEFT approximation. The electroweak corrections 
are also known to NLO 
\cite{Aglietti:2004nj,
Actis:2008ug} which amount to a positive correction 
of about $5\%$ compared to the NNLO QCD. 
Higgs production through bottom annihilation, 
despite being the subdominant channel,  has also 
been studied extensively in the literature. 
Due to the availability of third order form factor \cite{Gehrmann:2014vha}, the soft 
distributions \cite{Anastasiou:2014vaa},
and the relevant splitting functions \cite{Vogt:2004mw,Moch:2004pa},
the inclusive cross-section is known up to N3LO 
\cite{Dicus:1998hs,
Balazs:1998sb,
Maltoni:2003pn,
Harlander:2003ai,
Duhr:2019kwi} 
in the 5FS where the residual scale uncertainties 
are found to be around 5\%  and to NLO 
\cite{Dittmaier:2003ej,
Dawson:2003kb,
Wiesemann:2014ioa} 
in the 4FS. There have been several studies to 
combine the 5FS and 4FS prediction through 
different matching prescriptions 
\cite{Aivazis:1993pi,
Cacciari:1998it,
Forte:2010ta,
Harlander:2011aa,
Bonvini:2015pxa,
Forte:2015hba,
Bonvini:2016fgf,
Forte:2016sja,
Duhr:2020kzd}.
Further a complete N3LL resummation is performed in \cite{Ajjath:2019neu,Das:2022zie} where the 
scale uncertainty at this order reduces to about $4.9\%$.
The pure QED and mixed QCD-QED effects are studied 
in \cite{Ajjath:2019ixh} where the corrections are found to be below $0.03\%$ of the LO.

%
Differential measurement like the 
rapidity distribution of the Higgs boson 
is important to understand Higgs interaction 
within the SM. Similar to the total production 
cross-section, rapidity is also inclusive to 
extra radiations. 
While this sheds light 
on the spin of the particle itself, it is also useful 
to constrain the parton distribution functions (PDFs). 
In particular, the region 
with large momentum fraction is not 
well-constrained where the resummed results 
could play an important role.
Rapidity distribution has been known to NNLO 
for the Higgs production through gluon fusion 
\cite{Anastasiou:2004xq,Anastasiou:2005qj},
as well as in bottom quark annihilation 
\cite{Buhler:2012ytl,Mondini:2021nck}.
The accuracy is further extended beyond the NNLO 
level by studying the threshold contributions 
up to the third order 
\cite{Ravindran:2006bu,Ahmed:2014uya}
in gluon fusion as well in to bottom annihilation 
\cite{Ahmed:2014era}. 
It was observed there that 
the threshold contributions play a prominent role 
in the rapidity distributions for these processes.
Recently even complete N3LO corrections are also 
obtained for Higgs \cite{Cieri:2018oms}
and Drell-Yan (DY) \cite{Chen:2021vtu} rapidities 
using $q_T$ subtractions 
\cite{Catani:2007vq,Catani:2009sm,Catani:2010en,Catani:2011qz}
where the corrections are 
shown to be around $3\%$ and  $-2\%$ respectively over
NNLO in the central rapidity range. 
However, unlike the DY case, the uncertainty band 
for the Higgs rapidity distribution at N3LO 
show a nice perturbative convergence, reducing the 
scale uncertainty below $5\%$ and residing within the 
NNLO uncertainty band. Beyond NNLO, large threshold 
logarithms are also included at next-to-next-to-leading 
logarithmic (NNLL) accuracy \cite{Banerjee:2017cfc} 
in the gluon fusion channel.
Efforts \cite{Ajjath:2020lwb} are also made to 
resum partial subleading logarithms in the gluon fusion.

In this article we focus on the bottom induced 
rapidity distribution which is known to NNLO 
\cite{Buhler:2012ytl,Mondini:2021nck} for quite 
some time. We aim to improve this by including 
the threshold effects at NNLL and beyond.  
Within the traditional QCD resummation 
framework, a formalism 
\cite{Catani:1989ne,
Westmark:2017uig,
Banerjee:2017cfc,
Banerjee:2018vvb,
Banerjee:2018mkm,
Ahmed:2020amh} 
has been already developed to resum the large 
threshold logarithms in rapidity distribution for 
colorless particles. Originally the formalism was 
proposed for the $x_F$ distribution  in the seminal 
work \cite{Catani:1989ne} by Catani and Trentadue.
Later it was extended for rapidity\footnote{Note 
that the threshold behavior is same for both $x_F$ 
and rapidity.} 
following the framework developed in 
\cite{Ravindran:2006bu,Ravindran:2007sv}.
The idea is to identify proper scaling variables 
$z_1,z_2$ corresponding to partonic threshold ($z$)
and rapidity ($y_p$).  
One then resums large rapidity logarithms by
resumming these scaling variables simultaneously going 
to the threshold limit $z_1, z_2 \to 1$.
This was termed \cite{Banerjee:2017cfc,Banerjee:2018vvb,
Banerjee:2018mkm} as the Mellin-Mellin (M-M)
approach as the resummation is performed in the 
double Mellin space corresponding to $z_1, z_2$.
Essentially in this approach one resums all the 
double singular terms arising from the delta function 
$\delta(\bar{z}_i)$  and plus-distributions 
$\left[\frac{\ln^n\bar{z}_i}{\bar{z}_i} \right]_+$
where $\bar{z}_i = 1-z_i$.
This is also consistent with the 
generalized threshold resummation approach 
\cite{Lustermans:2019cau} employing  
soft-collinear effective theory (SCET) in the 
double singular limit. A recent comparison for different 
approaches up to the 
next-to-leading power (beyond double soft) level 
can be found in \cite{Bonvini:2023mfj}.

In this article we follow the standard M-M approach
at the leading power within the traditional QCD resummation framework
and study the impact of these threshold logarithms 
for the Higgs boson rapidity distribution 
in the bottom quark annihilation channel. 
We organize the paper as follows:
in \sect{sec:THEORY} we lay out the theoretical 
framework for the resummation of rapidity 
distributions in the M-M approach, in 
\sect{sec:NUMERICS} we present the results 
relevant at the 13 TeV LHC, and finally we 
conclude in \sect{sec:CONCLUSION} and collect 
all the analytical results required upto N3LL 
in the appendices (\ref{App:ANOMALOUS-DIMENSIONS} - \ref{App:SV-3LOOP}).
\section{Theoretical Framework} \label{sec:THEORY}
The effective Lagrangian for the interaction of a 
scalar Higgs boson with the bottom quark is given as,
\begin{align}\label{eq:LAGRANGIAN}
{\cal L}_{\rm int}^{\rm (S)} 
= -\lambda~ \bar{\psi}_{b}(x)\psi_{b}(x)\phi(x) \,,
\end{align}
where $\psi_{b}(x)$ 
and $\phi(x)$ are the bottom quark field 
and the Higgs field respectively.
Here $\lambda$ is the Yukawa interaction which is 
given by $\lambda = m_b/v$, with $v$ being 
the vacuum expectation value (VEV), and 
$m_b$ being mass of the bottom quark. 
We follow the 5FS where
we use non-zero mass of the bottom quark only 
in the Yukawa coupling, elsewhere it is treated as a
massless quark. 
The rapidity distribution of the Higgs boson at 
proton-proton collider takes the following form, 
\begin{align}\label{eq:RAPIDITY-MASTER}
        {\df\sigma(\tau, y)\over \df y } 
        =
        \sum_{i,j=q,\bar{q},g}
        &\int_{0}^1 {\df  x_1}
        \int_{0}^1 {\df  x_2}~
        f_i\left({x_1},\muf\right)  
        f_j\left({x_2},\muf\right)
\nn & \times 
        \int_{0}^1 {\df  z_1}
        \int_{0}^1 {\df z_2}~
        \delta(x_1^0 - x_1 z_1)
        \delta(x_2^0 - x_2 z_2)~
\widehat{\sigma}_{d,ij} (z_1,z_2,\muf,\mur)\,.
\end{align} 
Here $\tau=m_H^2/S = x_1^0 x_2^0 $ 
with $S$ being the hadronic center of mass energy.
The hadronic rapidity is defined as 
$y = \frac{1}{2}\ln \left( x_1^0/x_2^0\right)$.
The rapidity-dependent partonic coefficient function 
($\widehat{\sigma}_{d,ij}$) can be decomposed in 
terms of singular SV piece consisting of 
plus-distributions and delta function in 
partonic threshold variables ($z_1,z_2$)
and  non-singular or regular piece as,
\begin{align}\label{eq:partonic-decompose}
\widehat{\sigma}_{d,ij}(z_1,z_2,\muf,\mur)
=
\sigma_{0}(\mur) 
\Big( 
	\Delta_{d,ij}^{\rm sv}\left(z_1,z_2,\muf,\mur\right) 
	+ \Delta_{d,ij}^{\rm reg}\left(z_1,z_2,\muf,\mur\right)
\Big) \,.
\end{align}
The singular SV ($\Delta_{d,ij}^{\rm sv}$)
part gets contributions 
only from the diagonal channel \text{i.e.}\
in the present case $i,j = b, \bar{b}$ for the 
SV part \cite{Ahmed:2014cha}. 
On the other hand, the regular terms 
($\Delta_{d,ij}^{\rm reg}$)
are subdominant at the threshold 
and gets contributions from all partonic channels.
The overall Born normalization factor 
$\sigma_{0}(\mur) $ takes the following form,
\begin{align}
\sigma_{0}(\mur) 
=
\frac{\pi \lam^{2}(\mur) \tau}{6 m_H^2}
\equiv
\frac{\pi m_b^{2}(\mur) \tau}{6 m_H^2 v^2} \,.
\end{align}
Note that the $\mur$ dependence in the Born factor 
above comes only through the running of bottom mass 
($m_b(\mur)$) or equivalently the Yukawa 
($\lambda(\mur)$). 
The Yukawa running is performed using the 
mass anomalous dimensions $\gamma_m$ which we 
collect in the appendix\app{App:YUKAWA-RUNNING}
up to the third order.

Resummation is conveniently performed in the 
Mellin space where the double Mellin transformation 
is performed taking Mellin transformation 
on both partonic threshold variables $z_1, z_2$
in the following way (suppressing the $\mur,\muf$ 
dependence),
\begin{align}\label{eq:mellin-partonic}
        \Delta_{d,b\bar{b}}(N_1,N_2) 
        =& 
        \int_0^1  \df z_1   z_1^{N_1-1}   
        \int_0^1  \df z_2   z_2^{N_2-1}   
        \Delta^{\rm sv}_{d,b\bar{b}} (z_1,z_2) \,.
\end{align}
Here $N_i$ is the Mellin variable corresponding
to partonic threshold $z_i$.
The singular SV contribution can be resummed 
through the integral form in terms of 
universal cusp anomalous 
dimensions $A^b$ and rapidity-dependent threshold 
non-cusp anomalous dimension $D_d^b$ as well as 
process-dependent coefficients $g_{0}^{'b}$. 
In the double-Mellin space this can be written in the 
following integral form,
\begin{align}\label{eq:mellin-integral-form}
\widetilde{\Delta}_{d,b\bar{b}}(N_1,N_2) 
= &
g^{'b}_{_{d,0}}(\as) \exp
\Bigg(  
        \int_0^1 \df z_1 z_1^{N_1-1}   
        \int_0^1 \df z_2 z_2^{N_2-1}  
\nn&
\times \bigg(
        \delta(\bar{z}_2)
        \Bigg[
        \frac{1}{\bar{z}_1} 
        \Bigg\{ \int_{\muf^2}^{M_H^2 \bar{z}_1}
                {\df \eta^2 \over \eta^2}~ 
                A^b\left(a_s(\eta^2)\right) 
                +D^b_d\left(a_s(M_H^2~\bar{z}_1)\right) 
        \Bigg\}
        \Bigg]_+  
\nn&
        +
        \Bigg[
                \frac{1}{2\bar{z}_1\bar{z}_2} 
        \Bigg\{A^b(a_s(z_{12})) 
        + \frac{\df D^b_{d}(a_s(z_{12}))}{\df\ln z_{12}} 
        \Bigg\}
        \Bigg]_+
        + (z_1 \leftrightarrow z_2)
\bigg)
\Bigg)\,,
\end{align}
where $z_{12} = M_H^2 z_1 z_2$ and $\bar{z}_i = 1- z_i$.
Expansion of the above expression at a fixed order 
in strong coupling will reproduce the Mellin space 
fixed order results corresponding to \eq{eq:mellin-partonic}.
Performing the Mellin integration above will 
produce some non-logarithmic constants in double 
Mellin space which can be 
combined with the process-dependent prefactor 
($g^{'b}_{_{d,0}}$) and one can finally organize  
the large threshold logarithms in the Mellin space 
in the following form,
\begin{align}\label{eq:RESUM-EXPRESSION}
        \widetilde{\Delta}_{d,b\bar{b}}(N_1,N_2)
= 
g^b_{_{d,0}}(\as,\mu_r,\mu_f) 
\exp 
\Bigg( 
        G^b_d(\as,\w,\mur, \muf)   
\Bigg)\,.
\end{align}
The non-logarithmic coefficient 
($g^b_{_{d,0}}$) has the following perturbative 
expansion 
\begin{align}
        g^b_{_{d,0}}
        = 
        1 + 
        \sum_{i=1}^{\infty} 
        \as^i ~
        g^{b,(i)}_{_{d,0}} \,.
\end{align}
To achieve N3LL accuracy, they are needed 
up to the third order in strong coupling. 
We have obtained these using the third order 
$b\bar{b} H$ form factor \cite{Gehrmann:2014vha} 
and third order soft distribution 
function \cite{Anastasiou:2014vaa}, and we present these
in appendix\app{App:g0}.
The exponent is universal and resums the large 
logarithms (which now appear in $N_i \to \infty $ 
limit) to all orders. It can be expanded 
in the strong coupling and the inclusion of 
successive terms defines the resummed 
order,\footnote{Note that 
in the threshold region $\ln(\Nbar1 \Nbar2) \sim 1/\as $.}
\begin{align}\label{eq:resum-exponent}
G^b_d(\as,\w)
=
g^b_{d,1}(\w) \ln(\Nbar1 \Nbar2) 
+ \sum_{i=0}^\infty a_S^i g^b_{d,i+2}(\w)\,,  
\end{align}
where 
$\w =  a_s \beta_0 \ln(\Nbar1 \Nbar2)$, 
with $\Nbari = e^{\gamma_E} N_i, i=1,2$.
These process-independent resummed exponent are 
same as the quark-initiated Drell-Yan process and
up to N3LL accuracy these can be found e.g. in 
\cite{Das:2023bfi}.

The resummed expression in \eq{eq:RESUM-EXPRESSION} only resums the leading 
singular terms which appear through the large logarithms of $N_1$ and $N_2$
in Mellin space. This lacks the subleading regular 
pieces which can be included through the available 
fixed order results to improve the resummed predictions.
However, one can not simply add them as the 
fixed order also contains the same logarithmic 
contributions up to a certain order which are already
taken into account in the resummed expression.
Therefore, a matching procedure has to be invoked removing 
these logarithms which also appear in the fixed order.
This is done through the following all order
matched expression,
\begin{align}\label{eq:master-matched}
\frac{\df\sigma^{\text{res}} }{ \df y } 
=  
\frac{\df \sigma^{\text{f.o.}} }{ \df y } 
+& \sigma_0(\mur) 
\sum_{k,l=b,\bar{b}}
\int_{c_{1} - i\infty}^{c_1 + i\infty} 
        \frac{d N_{1}}{2\pi i}
\int_{c_{2} - i\infty}^{c_2 + i\infty} 
        \frac{d N_{2}}{2\pi i} 
e^{y(N_{2}-N_{1})}
\left(\sqrt{\tau}\right)^{-N_{1}-N_{2}} 
\nn&
\times
\widetilde f_{k}(N_{1}) ~
\widetilde f_{l}(N_{2}) ~
\Big[ 
        \widetilde{\Delta}_{d,b\bar{b}}(N_1,N_2) 
        -\widetilde{\Delta}^{\rm f.o.}_{d,b\bar{b}}(N_1,N_2) 
\Big] \,.
\end{align}
The first term on the right-hand side of the equality 
is the fixed order contribution containing singular
and regular contribution up to a fixed order in 
strong coupling. The first term inside the 
square bracket organizes the resummed series 
up to a certain logarithmic accuracy provided by 
the knowledge of cusp and rapidity anomalous dimensions
as well as the process-dependent $g_{d,0}^b$ 
coefficients.
The symbol `\text{f.o.}'\ in the second term 
inside the square bracket   
means the function is 
truncated to a fixed order in order to avoid 
double counting of singular terms already present in 
the fixed order 
($\frac{\df \sigma^{\text{f.o.}} }{ \df y } $). 
Here $\widetilde f_j(N_i) \equiv \int_0^1 dz_i z_i^{N_i-1}f_j(z_i)$
are the PDFs in the Mellin space. 
In practice, we use the $x$-space PDF through the 
\texttt{LHAPDF6} \cite{Buckley:2014ana}
interface using the derivatives of PDF as described in 
\cite{Kulesza:2002rh}.

The Mellin inversion in \eq{eq:master-matched} is not 
straightforward as the resummed expression diverges 
when $\w = 1$. This corresponds to the Landau pole 
where strong coupling diverges. The perturbative 
formalism thus break down in this region. One way to 
proceed is by choosing the contour of the Mellin 
inversion according to the 
\textit{minimal prescription} (MP) \cite{Catani:1996yz}. 
The basic idea is to choose the contour in 
such a way that all the poles remain at the left of the 
contour except for the Landau pole which remains 
far right of the contour. In double Mellin space 
this is little involved as the Landau pole is now 
a function of two Mellin variables. Typically, one 
needs to project the complex integration on real 
variables $r_i$ and chose the contour accordingly.
One can still fix the contour of one of the Mellin 
variable ($N_1 = c_1 + r_1 \exp(i\phi_1)$) 
according to MP. Once one fixes the contour for $N_1$
through $c_1$ and $\phi_1$,
the Landau pole is not 
constrained anymore on the real axis of the 
second Mellin 
variable ($N_2 = c_2 + r_2 \exp(i\phi_2)$) and in fact, 
it now depends on the first Mellin 
variable ($N_1$). In order to satisfy the MP, 
a reasonable choice \cite{Westmark:2017uig,Das:2023bfi} 
of the contour for the second 
Mellin variable as $\phi_2 = 
\max\left(
        \pi/2
        -1/2\arg(\frac{1}{N_1} \exp(\frac{1}{\as \bt0} - 2 \gamma_E))
        ,3\pi/4 \right)
$.

\section{Numerical Results}\label{sec:NUMERICS}
With the setup introduced in the previous section, 
we now focus on the numerical impact of the threshold 
logarithms in the rapidity distribution.
We focus on $13$ TeV LHC with \texttt{CT14} as our 
default PDF choice which is used through the 
\texttt{LHAPDF6} \cite{Buckley:2014ana} interface.\footnote{
The fixed order results for $b\bar{b}H$ rapidity at NNLO are 
available with \texttt{CT14} PDF set from \cite{Mondini:2021nck}.}
The fixed order results are obtained from  \cite{Mondini:2021nck}
using the N-jettiness slicing method \cite{Boughezal:2015dva,Gaunt:2015pea}
as implemented in MCFM \cite{Campbell:1999ah,Campbell:2015qma,Boughezal:2016wmq}. 
The strong coupling is taken  
through the corresponding PDF sets.
At the third order we evolve the strong coupling 
using four-loop QCD beta function 
\cite{Gross:1973id,Politzer:1973fx,Caswell:1974gg,Jones:1974mm,Egorian:1978zx,Tarasov:1980au,Larin:1993tp,vanRitbergen:1997va,Czakon:2004bu,Baikov:2016tgj,Herzog:2017ohr,Luthe:2017ttg} 
for which we set the initial condition as 
$\als(m_Z) = 0.118$ where $m_Z = 91.1876$ GeV 
is the $Z$ boson mass. We set the Higgs mass to 
be $m_H = 125$ GeV. 
The central scale choice for this process is 
taken as $(\mur^c,\muf^c) = (1, 1/4)m_H$ GeV.
The choice of the low $\muf^c$ scale  is done
following the observation in \cite{Maltoni:2003pn} 
to minimize the effect of large collinear 
logarithms which appear at $\muf^c \sim m_H/4$. 
The Yukawa coupling is also evolved through renormalisation group equation 
using 4-loop mass anomalous dimensions  
with bottom mass $m_b(m_b) = 4.18$ GeV.
The required mass anomalous dimensions are 
collected in the appendix\app{App:YUKAWA-RUNNING}.
To have an estimation of the residual scale 
uncertainties we follow the standard seven-point 
scale variations around the central scale choice 
stated above, with the restriction 
$1/2\leq (\mur/\mur^c)/(\muf/\muf^c) \leq 2$. 
This amounts to seven configurations for the scale $(\mur,\muf)$.
For each bin, the uncertainty envelope is obtained 
by considering the maximum and minimum deviations 
from the central scale choice. Note that the scale variation
also includes the scale dependence as arising from the 
Yukawa running in the $\overline{\text{MS}}$ scheme.
The double Mellin inversion in 
\eq{eq:master-matched} is performed with an 
in house code which we also interface to 
\texttt{LHAPDF6} as well as to 
\texttt{Cuba} \cite{Hahn:2004fe,Hahn:2014fua} for the final integration. 
Accordingly we chose the contour in \eq{eq:master-matched} 
as $c_1=c_2 =1.9$ and $\phi_1 = 3\pi/4$ and 
$\phi_2$ given in the previous section.

We further define the following perturbative 
quantities in order to assess the higher 
order effects. 
We define the ratios $K$-factor and $R$-factor 
\cite{Das:2019bxi,Das:2020gie,Das:2020pzo} 
corresponding to the fixed order and resummed 
order respectively as,
\begin{align}\label{eq:K-R-FACTORS}
        K_{ij} = 
        {\left[\frac{\df \sigma}{\df y}\right]_{{\rm N}i{\rm LO}} }\bigg/
        {\left[\frac{\df \sigma}{\df y}\right]_{{\rm N}j{\rm LO}} }\,,
        R_{ij} = 
        {\left[\frac{\df \sigma}{\df y}\right]_{{\rm N}i{\rm LO}+ {\rm N}i{\rm LL}} }\bigg/
        {\left[\frac{\df \sigma}{\df y}\right]_{{\rm N}j{\rm LO}+ {\rm N}j{\rm LL}} }\,.
\end{align}
Further we define the $RF_{ij}$ rations to estimate the 
resummed contributions over the fixed orders as,
\begin{align}\label{eq:R-F-FACTORS}
        RF_{ij} = 
        {\left[\frac{\df \sigma}{\df y}\right]_{{\rm N}i{\rm LO}+ {\rm N}i{\rm LL}} }\bigg/
        {\left[\frac{\df \sigma}{\df y}\right]_{{\rm N}j{\rm LO}} }\,.
\end{align}
For such ratios, we calculate the correlated error 
by taking both the numerator and the denominator 
at the same scale and obtain the seven-point 
uncertainties from seven such ratios.
In order to shorten the notation, we denote 
+LL, +NLL, +NNLL to indicate the complete matched 
resummed results, meaning the resummed corrections 
are matched to the FO results according to 
\eq{eq:master-matched}. At the third order similar 
notation is adopted where the complete resummed N3LL 
resuls are matched to the N3LOsv results and are denoted 
as +N3LLsv.

\begin{figure}[ht!]
        \centering{
\includegraphics[width=7.4cm,height=5.6cm]{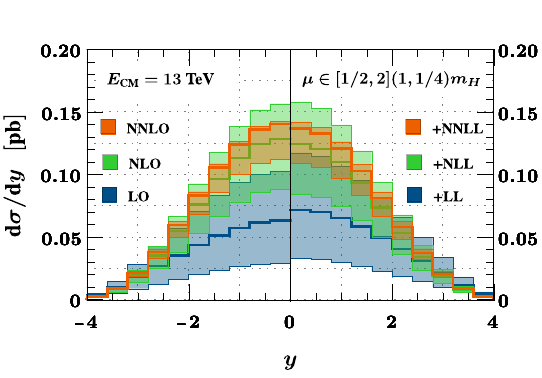}
\hspace{0.05cm}
\includegraphics[width=7.4cm,height=5.6cm]{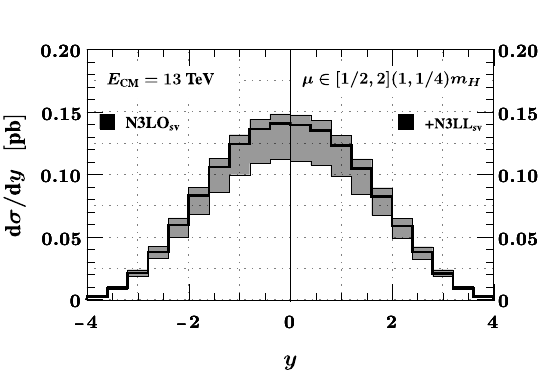}
}
\caption{Higgs rapidity distribution in bottom 
annihilation for $13$ TeV LHC. The left figure compares 
different fixed order and matched resummed orders.
The notation +N$k$LL indicates that the matched resummed 
results are obtained by matching with the corresponding 
FO results.
The right figure compares the SV results at the third order 
against the matched result at the same order where the 
matching is done with corresponding SV results.}
\label{fig:BBH-RAPIDITY}
\end{figure} 
In \fig{fig:BBH-RAPIDITY}, we present the 
rapidity distribution of the Higgs boson 
in bottom quark annihilation up to 
NNLO (left panel) and to +NNLL (right panel).
The asymmetric band is obtained by taking the 
envelope of maximum(minimum) deviation from the 
central scale according to the seven-point scale 
variation. 
On the fixed order side, the NLO 
gets a correction of similar size to LO, 
which gets further increased at NNLO by $9.3\%$ 
compared to NLO in the central rapidity region (at $y=0-1.6$).
In the higher rapidity region (at $y=2-4$), the behavior 
is different where NLO gets negative correction 
up to $50\%$ compared to LO, whereas on the 
other hand, the NNLO gets positive correction 
up to $27\%$ of NLO.
The corresponding scale uncertainties are 
respectively \uncert[+62.9]{-53.9} at LO, 
\uncert[+22.0]{-31.1} at NLO, and 
\uncert[+1.7]{-20.0} at NNLO in the central rapidity 
region. In the resummed case, 
the convergence is faster with +NLL getting a 
correction $73\%$ compared to +LL
whereas +NNLL gets a further increment of 
$10.1\%$ compared to +NLL in the central region.
Corresponding corrections at the 
higher rapidity ($y=3.2$) region are $-31.7\%$
and $8.6\%$ respectively. 
The scale uncertainty does not improve compared to 
the FO, a behavior which is also seen in the 
neutral and charged DY productions 
\cite{Das:2023bfi}. In the central region, 
the asymmetric scale uncertainties are 
\uncert[+63.4]{-54.0} at +LL, 
\uncert[+27.1]{-32.2} at +NLL, 
\uncert[+3.8]{-20.4} at +NNLL respectively.
At $y=3.2$ the corresponding uncertainties are 
\uncert[+52.9]{-50.4} at +LL, 
\uncert[+19.8]{-27.6} at +NLL, 
\uncert[+11.9]{-5.4} at +NNLL respectively.
We notice that while +LL gets a positive 
contribution ranging from $13.7\% $ to $ 22.8\%$ 
over LO, the +NLL gets $-3.2\%$ to $-2.3\%$ 
corrections over NLO
and +NNLL gets relatively flat correction 
ranging $-2.4\%$ to $ - 3.1\%$ 
going from central rapidity to higher rapidity.
On the right panel of \fig{fig:BBH-RAPIDITY}, 
we present the new third order SV results which 
is further matched with the third order resummed 
results. While SV results at the third order 
gets a increment up to  $0.5\%$ compared to the NNLO,
the scale uncertainty is not improved. 
The third order matched results however 
gets a flat correction of about $1.8\%$
throughout the rapidity region 
compared to +NNLL with the scale uncertainties 
are at the similar level as the fixed order. 

\begin{figure}[ht!]
	\centering{
		\includegraphics[width=7.4cm,height=5.6cm]{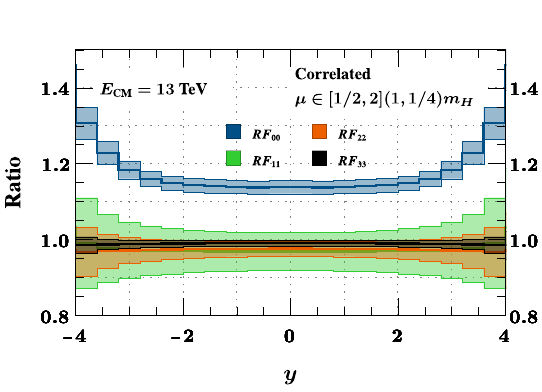}
	}
	\caption{$RF$ factor (as defined in 
		\eq{eq:R-F-FACTORS})
		along with correlated 
		errors up to the third order. 
		For the third order, resummed N3LL results 
		are matched to the N3LOsv results.}
	\label{fig:BBH-RF-RATIOS}
\end{figure}
It is also possible to match the resummed result in a 
multiplicative way instead of the additive matching in 
\eq{eq:master-matched}.
In order to estimate the effect of such a procedure,
we followed the prescription\footnote{See Eq. (4.3) of 
\cite{Bizon:2018foh}.}  
as presented in \cite{Bizon:2018foh}. 
We find that 
the +NLL gets an increment of around $84\%$ at $y=0$ 
compared to +LL whereas at +NNLL the correction is less 
that $1\%$ of +NLL. Similarly, at $y=3.2$, the corrections
are $-31\%$ and $5\%$ respectively. 
Thus, we observe a faster convergence for multiplicative
matching compared to the additive matching.
The corresponding asymmetric scale uncertainties at 
$y=3.2$ are
\uncert[+53.9]{-50.8} at +LL, 
\uncert[+25.0]{-33.8} at +NLL, 
\uncert[+6.4]{-2.9} at +NNLL respectively.
Up to +NLL, the uncertainties remains similar compared 
to the additive matching, at +NNLL, the uncertainties gets 
better at higher rapidities. On the other hand we see 
similar scale uncertainties as the additive case at $y=0$  
at +NNLL level.
In the rest of the article we follow the additive 
matching as provided by \eq{eq:master-matched}.

\begin{figure}[ht!]
        \centering{
\includegraphics[width=7.4cm,height=5.6cm]{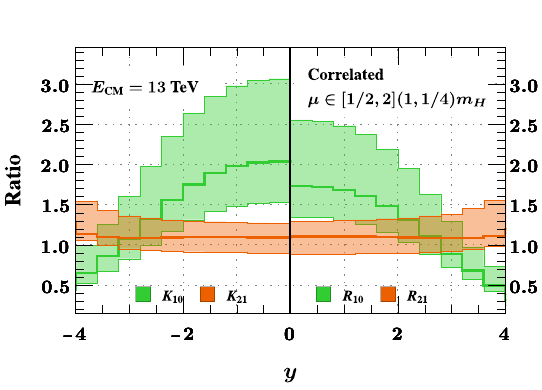}
\hspace{0.05cm}
\includegraphics[width=7.4cm,height=5.6cm]{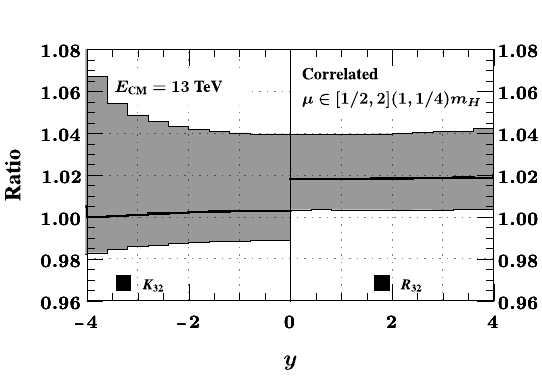}
}
\caption{$K, R$ factors (as defined in 
\eq{eq:K-R-FACTORS})
along with correlated 
errors upto the second order (left) and 
the same at the third order (right). The third 
order results are up to SV accuracy and same for 
matched.}
\label{fig:BBH-SRF-RATIOS}
\end{figure}

A better way to visualize the higher order 
effects is through the 
ratios {\it viz.} the $K$, $R$, and $RF$ factors as 
defined in \eq{eq:K-R-FACTORS}-\eq{eq:R-F-FACTORS} 
which are 
presented in  \fig{fig:BBH-SRF-RATIOS} and \fig{fig:BBH-RF-RATIOS}.
In the left panel we show these ratios 
up to the second order. It is clear that 
NLO (or +NLL) corrections shapes the rapidity 
distributions very well,
whereas the corrections from NNLO (or +NNLL)
are rather flat over a large rapidity region. 
Compared to the $K_{21}$ factor we observe slight 
increment of the central scale 
in the case of $R_{21}$ in the higher 
rapidity region.
On the right panel of \fig{fig:BBH-SRF-RATIOS},
we present the these ratios at the third order. 
Again we observe a relatively flat QCD correction 
over the large rapidity range at N3LOsv amounting 
to about $8\%$ uncertainty at the higher 
rapidities. The matched result becomes almost flat 
even in the higher rapidity region and the 
correlated scale uncertainty reduces to 
$4\%$.

\begin{figure}[ht!]
        \centering{
\includegraphics[width=7.4cm,height=5.6cm]{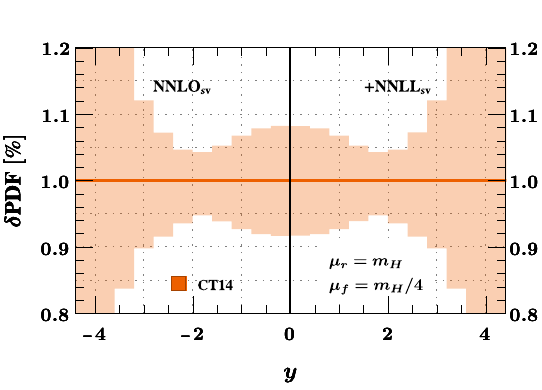}
}
\caption{Effects of matched NNLL results to NNLOsv are 
presented and compared against the NNLOsv for \texttt{CT14} 
PDF at $13$ TeV LHC.}
\label{fig:BBH-PDF-UNCERTAINTY}
\end{figure}
To estimate the intrinsic PDF uncertainty, we 
define the quantity 
$\delta {\rm PDF} = 1\pm\delta [\frac{\df \sigma}{\df y}]/[\frac{\df \sigma}{\df y}]_0 \times 100\% $, 
where the numerator $\delta [\frac{\df \sigma}{\df y}]$ is 
the intrinsic PDF uncertainty and the denominator 
$[\frac{\df \sigma}{\df y}]_0$ is the central prediction.
We study only the SV part of the fixed order 
which might be improved by resummed results.
This is shown in \fig{fig:BBH-PDF-UNCERTAINTY}. 
Notice that the central predictions are different for 
FO and resummed cases. Indeed, we observe a 
$-2.5\%$ to $-2.1\%$ change in the central predictions 
for resummed case compared to FO SV 
while going from $y=0$ to $y=4$. 
The PDF uncertainties at the second order (SV) 
results are about $8\%$ in the central rapidity ($y=0$) 
which gets reduced to about $4.5\%$ at $y=-2$.
At the higher rapidity region the uncertainties
further increases. This is a known behavior which 
is also observed in the case of DY rapidities 
\cite{Das:2023bfi}. The matched resummed results 
show a slight improvement ( below $0.1\%$)  
over the fixed order PDF uncertainty in all rapidity region.

\section{Conclusions}\label{sec:CONCLUSION}
At the LHC, soft gluons play an important role 
in predicting observables correctly in different 
phase space corners. Particularly in the threshold 
region their contribution dominates and hence to 
have a reliable prediction one needs to resum them 
and match them to the available fixed order to 
better predict an observable. For rapidity 
distribution in the threshold region  both 
threshold variables corresponding to partonic 
threshold and rapidity become equally 
important and one needs to resum both of them 
consistently order by order in the perturbation theory. 
In QCD, this is achieved by the M-M approach, and 
we exploit this to 
resum large threshold logarithms to NNLL accuracy 
matched to available NNLO result. The threshold 
effects at fixed order amount to $-3.2\%$ 
enhancement over 
the NLO distribution over a large range of rapidity. 
On the other hand, the resummed corrections at NNLO+NNLL 
amount to $-2.5\%$ enhancement over the fixed 
order NNLO. 
In general, we find a better perturbative convergence
in the resummed spectrum which combines merit of both 
resummed logarithms and non-singular 
contributions from fixed order which are not captured 
in purely resummed predictions. While the third order 
analytic ingredients presented in this article will 
be useful to match to the third order fixed order results once 
it will be available, we present the 
first predictions at the third order coming from the 
dominant threshold logarithms and also match them to 
N3LL. We observe a relatively flat correction in the 
third order matched results for all rapidities
with reduced scale uncertainties to below $4\%$ 
particularly in 
the higher rapidities, a region dominated by 
the large threshold logarithms. 
Our resummed results can be useful in 
constraining the bottom quark PDFs at the large 
momentum fraction.
\section*{Acknowledgments}
G.D.\ is grateful to C.\ Williams for providing 
the NNLO results.
G.D.\ also thanks V.\ Ravindran for encouraging to work on 
this project.
Part of the analytical computation has been 
performed using symbolic manipulation system 
F{\sc orm} \cite{Vermaseren:2000nd,Ruijl:2017dtg}.
Simulations were performed with computing resources granted by
RWTH Aachen University under project 
{\tt rwth1298}.
The research of G.D. was supported by the
Deutsche Forschungsgemeinschaft
(DFG, German Research Foundation)
under grant  396021762 - TRR 257
(\textit{Particle Physics Phenomenology 
after Higgs discovery.}).
\appendix
\section{Anomalous Dimensions}\label{App:ANOMALOUS-DIMENSIONS}
The cusp anomalous dimensions have the following 
perturbative expansion in strong coupling,
\begin{align}
        A^b = \sum_{i=1}^{\infty}
        \as^{i}~ A_{i}^b \,.
\end{align}
We collect here the coefficients up to fourth order
\cite{Henn:2019swt,Huber:2019fxe,vonManteuffel:2020vjv} 
needed at N3LL,  
\begin{align}
\begin{autobreak}
        A^b_1  =
        4 C_F \,, 
\end{autobreak}
\nn
\begin{autobreak}
        A^b_2  =
        8 C_F C_A 
        \bigg( 
                \frac{67}{18} 
                - \zeta_2 
        \bigg) 
        + 8 C_F n_f 
        \bigg( -\frac{5}{9} 
        \bigg) \,,
\end{autobreak}
\nn
\begin{autobreak}
        A^b_3  =
        16 C_F C_A^2 
        \bigg( 
        \frac{245}{24} 
        - \frac{67}{9} \zeta_2  
        + \frac{11}{6} \zeta_3
        + \frac{11}{5} \zeta_2^2 
        \bigg)
        + 16 C_F^2 n_f 
        \bigg( 
        - \frac{55}{24} 
        + 2 \zeta_3 
        \bigg)
        + 16 C_F C_A n_f 
        \bigg( 
        - \frac{209}{108} 
        + \frac{10}{9} \zeta_2 
        - \frac{7}{3} \zeta_3         
        \bigg)
        - 16 C_F n_f^2 
        \bigg(  
        \frac{1}{27} 
        \bigg) \,,
\end{autobreak}
\nn
\begin{autobreak}
        A_4^b = 
        C_{F}\nf^3    
        \bigg(
        - \frac{32}{81}
        + \frac{64}{27}  \z3 
        \bigg)
        + \Cf^2  \nf^2    \bigg( 
        \frac{2392}{81}
        - \frac{640}{9}  \z3
        + \frac{64}{5}  \z2^2 
        \bigg)
        + \Ca \Cf  \nf^2    \bigg( 
        \frac{923}{81}
        + \frac{2240}{27}  \z3
        - \frac{608}{81}  \z2
        - \frac{224}{15}  \z2^2 
        \bigg)
        + \Cf^3  \nf    \bigg( 
        \frac{572}{9}
        - 320  \z5
        + \frac{592}{3}  \z3 
        \bigg)
        + \Ca^2 \Cf \nf    \bigg(
        - \frac{24137}{81}
        + \frac{2096}{9}  \z5
        - \frac{23104}{27}  \z3
        + \frac{20320}{81}  \z2
        + \frac{448}{3}  \z2  \z3
        - \frac{352}{15}  \z2^2 
        \bigg)
        + \Ca  \Cf^2  \nf    \bigg(
        - \frac{34066}{81}
        + 160  \z5
        + \frac{3712}{9}  \z3
        + \frac{440}{3}  \z2
        - 128  \z2  \z3
        - \frac{352}{5}  \z2^2 
        \bigg)
        + \Ca^3 \Cf    \bigg( 
        \frac{84278}{81}
        - \frac{3608}{9}  \z5
        + \frac{20944}{27}  \z3
        - 16  \z3^2
        - \frac{88400}{81}  \z2
        - \frac{352}{3}  \z2  \z3
        + \frac{3608}{5}  \z2^2
        - \frac{20032}{105}  \z2^3 
        \bigg)
        + \nf  \dFF    \bigg(
        - \frac{1280}{3}  \z5
        - \frac{256}{3}  \z3
        + 256  \z2 \bigg)
        + \dFA     \bigg( \frac{3520}{3}  \z5
        + \frac{128}{3}  \z3
        - 384  \z3^2
        - 128  \z2
        - \frac{7936}{35}   \z2^3 \bigg)
        \,.
\end{autobreak}
\end{align}
The quartic casimirs appearing above are defined as
\begin{align}
\frac{d_{FA}^{(4)}}{N_F} &\equiv
\frac{d_F^{abcd}d_A^{abcd}}{n_c} 
=  \frac{(n_c^2-1)(n_c^2+6)}{48}
= \frac{5}{2},
\nn
\frac{d_{FF}^{(4)}}{N_F} &\equiv
\frac{d_F^{abcd}d_F^{abcd}}{n_c} 
= \frac{(n_c^2-1)(n_c^4 - 6 n_c^2 + 18)}{96n_c^3}
= \frac{5}{36},
\end{align}
with $N_F = n_c = 3$ for QCD.
The threshold non-cusp anomalous dimensions $D_d^b$ has the 
following perturbative expansion in the strong coupling,
\begin{align}
        D_{d}^b 
        = \sum_{i=1}^{\infty} \as^i~ D_{d,i}^b \,.
\end{align}
The coefficients are the same as the quark ones 
(see {e.g.} \cite{Das:2023bfi}).
Up to the N3LL accuracy, these are needed up to third 
order \cite{Ravindran:2006bu} which we collect below, 
\begin{align}
D^b_{d,1} 
=&~ 
0,
\nn
D^b_{d,2} 
=&~
\CF \NF   
\Bigg(
        \frac{112}{27}
        - \frac{8}{3} \z2
\Bigg)
+ \CA \CF   
\Bigg(
        - \frac{808}{27}
        + 28 \z3
        + \frac{44}{3} \z2
\Bigg),
\nn
D^b_{d,3} 
=&~
\CF \NF^2  
\Bigg(
        - \frac{1856}{729} 
        - \frac{32}{27} \z3 
        + \frac{160}{27} \z2 
\Bigg)
+ \CA \CF \NF   
\Bigg(
        \frac{62626}{729}
        + \frac{208}{15} \z2^2
        - \frac{536}{9} \z3
        - \frac{7760}{81} \z2
\Bigg)
\nn&
+ \CF^2 \NF   
\Bigg(
        \frac{1711}{27}
        - \frac{32}{5} \z2^2
        - \frac{304}{9} \z3
        - 8 \z2
\Bigg)
+ \CA^2 \CF   
\Bigg(
        - \frac{297029}{729}
        - \frac{616}{15} \z2^2
        - 192 \z5
\nn&
        + \frac{14264}{27} \z3
        + \frac{27752}{81} \z2
        - \frac{176}{3} \z2 \z3
\Bigg).
\end{align}
\section{Yukawa running} \label{App:YUKAWA-RUNNING}
The Higgs Yukawa coupling with bottom quark 
is given as $\lambda = m_b/v$.
Here $m_b(\mur)$ is the $\overbar{\text{ MS}}$
running mass of the bottom quark.
Thus, the running of Yukawa goes through the 
running of bottom mass as,
\begin{align}
        \mur^2 \frac{\df}{\df \mur^2} \lam(\mur) 
        =
        \gamma_m(a_S) \lam(\mur)\,.
\end{align}
The mass anomalous dimension ($\gamma_m$) has the following 
perturbative expansion:
\begin{align}
\gamma_m
= \sum_{i=0}^{\infty} \as^{i+1} \gamma_{m}^{(i)}\,,
\end{align}
where the coefficients are known up to four loops 
\cite{Tarasov:1982plg,
Larin:1993tq,
Alekseev:1994xp,
Chetyrkin:1997dh,
Czakon:2004bu}. 
We collect all the coefficients up to 
four-loop order,
\begin{align} 
\begin{autobreak} 
\gm0= 
\Cf    \bigg( 3 \bigg) \,,
\end{autobreak} 
\\ 
\begin{autobreak} 
\gm1= 
\Cf^2    \bigg( \frac{3}{2} \bigg)      
+ \Ca  \Cf    \bigg( \frac{97}{6} \bigg)      
+ \nf  \Cf    \bigg( 
- \frac{5}{3} \bigg) \,,
\end{autobreak} 
\\ 
\begin{autobreak} 
\gm2= 
\Cf^3    \bigg( \frac{129}{2} \bigg)      
+ \Ca  \Cf^2    \bigg( 
- \frac{129}{4} \bigg)      
+ \Ca^2  \Cf    \bigg( \frac{11413}{108} \bigg)      
+ \nf  \Cf^2    \bigg( 
- 23
+ 24  \z3 \bigg)      
+ \nf  \Ca  \Cf    \bigg( 
- \frac{278}{27}
- 24  \z3 \bigg)      
+ \nf^2  \Cf    \bigg( 
- \frac{35}{27} \bigg) \,,
\end{autobreak} 
\\ 
\begin{autobreak} 
\gm3= 
\dFAbN    \bigg( \frac{69383}{21}
- \frac{10560}{7}  \z5
+ \frac{16384}{21}  \z3 \bigg)      
+ \Cf^4    \bigg( 
- \frac{1261}{8}
- 336  \z3 \bigg)      
+ \Ca  \Cf^3    \bigg( \frac{15349}{12}
+ 316  \z3 \bigg)      
+ \Ca^2  \Cf^2    \bigg( \frac{182015}{252}
- \frac{2200}{7}  \z5
+ \frac{2480}{21}  \z3 \bigg)      
+ \nf  \Cf^3    \bigg( \frac{148}{3}
- 240  \z5
- 444  \z3 \bigg)      
+ \nf  \Ca  \Cf^2    \bigg( 
- \frac{13139}{54}
+ 40  \z5
+ 784  \z3
- \frac{264}{5}  \z2^2 \bigg)      
+ \nf  \Ca^2  \Cf    \bigg( 
- \frac{59843}{324}
+ 200  \z5
- \frac{1732}{3}  \z3
+ \frac{264}{5}  \z2^2 \bigg)      
+ \nf^2  \Cf^2    \bigg( \frac{76}{27}
- 40  \z3
+ \frac{48}{5}  \z2^2 \bigg)      
+ \nf^2  \Ca  \Cf    \bigg( \frac{671}{162}
+ 40  \z3
- \frac{48}{5}  \z2^2 \bigg)      
+ \nf^3  \Cf    \bigg( 
- \frac{83}{81}
+ \frac{16}{9}  \z3 \bigg) \,.
\end{autobreak} 
\end{align}
\section{The process dependent coefficient $g^b_{d,0}$}\label{App:g0}
Below we present the new process dependent 
coefficients ($g^{b}_{d,0{\rm i}}$) up to N3LL accuracy,
\begin{align} 
\begin{autobreak} 
\g01bbH = 
\CF    \bigg\{ 
- 4
+ 16  \z2
+ \bigg(-6\bigg)  \Lfr \bigg\} ,  
\end{autobreak} 
\\ 
\begin{autobreak} 
\g02bbH = 
\CF  \NF    \bigg\{ \frac{8}{9}
+ \frac{8}{9}  \z3
- \frac{40}{3}  \z2
+ \bigg(\frac{2}{3}
+ \frac{16}{3}  \z2\bigg)  \Lfr
+ \bigg(\frac{16}{3}  \z2\bigg)   \Lqr
+ \bigg(-2\bigg)  \Lfr^2 \bigg\}      
+ \CF^2    \bigg\{ 16
- 60  \z3
- 32  \z2
+ \frac{552}{5}  \z2^2
+ \bigg(21
- 48  \z3
- 72  \z2\bigg)  \Lfr
+ \bigg(48  \z3
- 24  \z2\bigg)  \Lqr
+ \bigg(18\bigg)  \Lfr^2 \bigg\}      
+ \CA  \CF    \bigg\{ \frac{166}{9}
+ \frac{280}{9}  \z3
+ \frac{256}{3}  \z2
- \frac{92}{5}  \z2^2
+ \bigg(
- 12
- 24  \z3
- \frac{88}{3}  \z2\bigg)  \Lqr
+ \bigg(
- \frac{17}{3}
+ 24  \z3
- \frac{88}{3}  \z2\bigg)  \Lfr
+ \bigg(11\bigg)  \Lfr^2 \bigg\} ,  
\end{autobreak} 
\\ 
\begin{autobreak} 
\g03bbH = 
\CF  \NF^2    \bigg\{ \frac{16}{27}
+ \frac{160}{81}  \z3
+ \frac{256}{27}  \z2
+ \frac{448}{135}  \z2^2
+ \bigg(
- \frac{8}{9}\bigg)   \Lfr^3
+ \bigg(\frac{4}{9}
+ \frac{32}{9}  \z2\bigg)  \Lfr^2
+ \bigg(\frac{34}{9}
+ \frac{32}{9}  \z3
- \frac{160}{27}  \z2\bigg)  \Lfr
+ \bigg(
- \frac{64}{27}  \z3
- \frac{320}{27}  \z2\bigg)  \Lqr
+ \bigg(\frac{32}{9}  \z2\bigg)  \Lqr^2 \bigg\}      
+ \CF^2  \NF    \bigg\{ 
- \frac{70}{9}
- \frac{608}{9}  \z5
+ \frac{8872}{27}  \z3
- \frac{3428}{27}  \z2
- \frac{256}{3}  \z2   \z3
- \frac{15688}{135}  \z2^2
+ \bigg(
- 4
- 32  \z3
- 48  \z2\bigg)  \Lfr^2
+ \bigg(\frac{8}{3}
- \frac{656}{3}  \z3
+ \frac{200}{3}  \z2
+ \frac{464}{5}  \z2^2\bigg)  \Lqr
+ \bigg(38
+ \frac{256}{3}  \z3
+ 56  \z2
+ \frac{272}{5}  \z2^2\bigg)   \Lfr
+ \bigg(32  \z3
- 16  \z2\bigg)  \Lqr^2
+ \bigg(
- 32  \z2\bigg)  \Lqrfr
+ \bigg(12\bigg)  \Lfr^3 \bigg\}      
+ \CF^3    \bigg\{ \frac{1078}{3}
+ 848  \z5
- 1188  \z3
+ 32  \z3^2
- \frac{166}{3}  \z2
- 544  \z2  \z3
- \frac{744}{5}  \z2^2
+ \frac{169504}{315}  \z2^3
+ \bigg(
- 113
+ 480  \z5
+ 416  \z3
+ 12  \z2
- 704  \z2  \z3
- \frac{1968}{5}  \z2^2\bigg)  \Lfr
+ \bigg(
- 100
- 480  \z5
- 56  \z3
+ 132  \z2
+ 704  \z2  \z3
- \frac{1344}{5}  \z2^2\bigg)  \Lqr
+ \bigg(
- 54
+ 288  \z3
+ 144  \z2\bigg)  \Lfr^2
+ \bigg(
- 288  \z3
+ 144  \z2\bigg)  \Lqrfr
+ \bigg(-36\bigg)  \Lfr^3 \bigg\}      
+ \CA  \CF  \NF    \bigg\{ 
- \frac{11540}{81}
- 8  \z5
- \frac{15944}{81}  \z3
- \frac{13040}{81}  \z2
+ \frac{880}{9}   \z2  \z3
+ \frac{184}{135}  \z2^2
+ \bigg(
- 40
- \frac{400}{9}  \z3
+ \frac{2672}{27}  \z2
- \frac{8}{5}  \z2^2\bigg)  \Lfr
+ \bigg(
- \frac{146}{9}
+ 16  \z3
- \frac{352}{9}  \z2\bigg)  \Lfr^2
+ \bigg(
- 8
- 16  \z3
- \frac{352}{9}  \z2\bigg)   \Lqr^2
+ \bigg(\frac{88}{9}\bigg)  \Lfr^3
+ \bigg(\frac{196}{3}
+ \frac{3440}{27}  \z3
+ \frac{4480}{27}  \z2
- \frac{344}{15}  \z2^2 \bigg)  \Lqr \bigg\}      
+ \CA  \CF^2    \bigg\{ 
- \frac{982}{3}
- \frac{3352}{9}  \z5
- \frac{11188}{27}  \z3
+ \frac{592}{3}  \z3^2
+ \frac{19658}{27}  \z2
+ \frac{2528}{3}  \z2  \z3
+ \frac{25676}{27}  \z2^2
- \frac{123632}{315}  \z2^3
+ \bigg(
- \frac{327}{2}
- 240  \z5
- \frac{2536}{3}  \z3
- 212  \z2
+ 352  \z2  \z3
- \frac{1136}{5}  \z2^2\bigg)  \Lfr
+ \bigg(1
+ 32   \z3
+ 264  \z2\bigg)  \Lfr^2
+ \bigg(72
+ 144  \z3
+ 176  \z2\bigg)  \Lqrfr
+ \bigg(\frac{388}{3}
+ 240  \z5
+ \frac{3296}{3}  \z3
- \frac{1748}{3}  \z2
- 352  \z2  \z3
- 472  \z2^2\bigg)  \Lqr
+ \bigg(
- 176  \z3
+ 88  \z2\bigg)  \Lqr^2
+ \bigg(-66\bigg)  \Lfr^3 \bigg\}      
+ \CA^2  \CF    \bigg\{ \frac{68990}{81}
- 84  \z5
+ \frac{42748}{81}  \z3
- \frac{400}{3}  \z3^2
+ \frac{39980}{81}  \z2
- \frac{7768}{9}  \z2  \z3
- \frac{25328}{135}  \z2^2
+ \frac{7088}{63}  \z2^3
+ \bigg(
- \frac{1180}{3}
+ 80  \z5
- \frac{15472}{27}  \z3
- \frac{12800}{27}  \z2
+ \frac{1964}{15}  \z2^2\bigg)  \Lqr
+ \bigg(
- \frac{242}{9}\bigg)  \Lfr^3
+ \bigg(44
+ 88  \z3
+ \frac{968}{9}  \z2\bigg)  \Lqr^2
+ \bigg(\frac{493}{9}
- 88  \z3
+ \frac{968}{9}  \z2\bigg)  \Lfr^2
+ \bigg(\frac{1657}{18}
- 80  \z5
+ \frac{3104}{9}  \z3
- \frac{8992}{27}  \z2
+ 4  \z2^2\bigg)  \Lfr \bigg\} \,.
\end{autobreak} 
\end{align}
\section{Soft-virtual coefficients in double Mellin space}\label{App:SV-3LOOP}
For completeness, here we collect all the singular SV 
coefficients \cite{Ahmed:2014cha} up to third order in the double Mellin 
space as defined in \eq{eq:mellin-partonic}. 
The perturbative expansion in Mellin 
space takes the following form,
\begin{align}
        {\Delta}_{d,b\bar{b}}(N_1,N_2)
        \equiv 
        \widetilde{\Delta}^{\rm f.o.}_{d,b\bar{b}}(N_1,N_2) 
        = 1 + \sum_{i=1}^{\infty}
        \as^i \Delta_{d,b\bar{b}}^{(i)}\,.
\end{align}
Defining 
$\LNoNtb \equiv \ln (\overbar{N}_1 \overbar{N}_2)$,
the coefficients up to third order take the form,
\begin{align} 
\begin{autobreak} 
\gSVN1 = 
\LNoNtb^2    \bigg\{ \bigg(2\bigg)  \CF \bigg\}      
+ \LNoNtb    \bigg\{ \bigg(\bigg(-4\bigg)  \Lqr
+ \bigg(4\bigg)  \Lfr\bigg)  \CF \bigg\}      
+\g01bbH\,,
\end{autobreak} 
\\ 
\begin{autobreak} 
\gSVN2 = 
\LNoNtb^4    \bigg\{ \bigg(2\bigg)  \CF^2 \bigg\}      
+ \LNoNtb^3    \bigg\{ \bigg(\bigg(-8\bigg)  \Lqr
+ \bigg(8\bigg)  \Lfr\bigg)  \CF^2
+ \bigg(
- \frac{4}{9}\bigg)  \CF  \NF
+ \bigg(\frac{22}{9}\bigg)  \CA   \CF \bigg\}      
+ \LNoNtb^2    \bigg\{ \bigg(
- 4  \z2
+ \bigg(
- \frac{22}{3}\bigg)  \Lqr
+ \frac{134}{9}\bigg)  \CA  \CF
+ \bigg(32  \z2
+ \bigg(-16\bigg)   \Lqrfr
+ \bigg(-12\bigg)  \Lfr
+ \bigg(8\bigg)  \Lqr^2
+ \bigg(8\bigg)  \Lfr^2
- 8\bigg)  \CF^2
+ \bigg(\bigg(\frac{4}{3}\bigg)  \Lqr
- \frac{20}{9}\bigg)  \CF  \NF \bigg\}      
+ \LNoNtb    \bigg\{ \bigg(
- 28  \z3
+ \bigg(
- 8  \z2
+ \frac{268}{9}\bigg)  \Lfr
+ \bigg(8  \z2
- \frac{268}{9}\bigg)  \Lqr
+ \bigg(
- \frac{22}{3}\bigg)  \Lfr^2
+ \bigg(\frac{22}{3}\bigg)  \Lqr^2
+ \frac{808}{27}\bigg)  \CA  \CF
+ \bigg(\bigg(
- 64  \z2
+ 16\bigg)  \Lqr
+ \bigg(64  \z2
- 16\bigg)  \Lfr
+ \bigg(-24\bigg)  \Lfr^2
+ \bigg(24\bigg)  \Lqrfr\bigg)  \CF^2
+ \bigg(\bigg(
- \frac{40}{9}\bigg)  \Lfr
+ \bigg(
- \frac{4}{3}\bigg)  \Lqr^2
+ \bigg(\frac{4}{3}\bigg)  \Lfr^2
+ \bigg(\frac{40}{9}\bigg)  \Lqr
- \frac{112}{27}\bigg)  \CF  \NF \bigg\}      
+\g02bbH\,,
\end{autobreak} 
\\ 
\begin{autobreak} 
\gSVN3 = 
\LNoNtb^6    \bigg\{ \bigg(\frac{4}{3}\bigg)  \CF^3 \bigg\}      
+ \LNoNtb^5    \bigg\{ \bigg(\bigg(-8\bigg)  \Lqr
+ \bigg(8\bigg)  \Lfr\bigg)  \CF^3
+ \bigg(
- \frac{8}{9}\bigg)  \CF^2  \NF
+ \bigg(\frac{44}{9}\bigg)  \CA   \CF^2 \bigg\}      
+ \LNoNtb^4    \bigg\{ \bigg(
- 8  \z2
+ \bigg(
- \frac{220}{9}\bigg)  \Lqr
+ \bigg(\frac{88}{9}\bigg)  \Lfr
+ \frac{268}{9}\bigg)  \CA  \CF^2
+ \bigg(32  \z2
+ \bigg(-32\bigg)  \Lqrfr
+ \bigg(-12\bigg)  \Lfr
+ \bigg(16\bigg)  \Lqr^2
+ \bigg(16\bigg)  \Lfr^2
- 8\bigg)   \CF^3
+ \bigg(\bigg(
- \frac{16}{9}\bigg)  \Lfr
+ \bigg(\frac{40}{9}\bigg)  \Lqr
- \frac{40}{9}\bigg)  \CF^2  \NF
+ \bigg(
- \frac{44}{27}\bigg)  \CA  \CF   \NF
+ \bigg(\frac{4}{27}\bigg)  \CF  \NF^2
+ \bigg(\frac{121}{27}\bigg)  \CA^2  \CF \bigg\}      
+ \LNoNtb^3    \bigg\{ \bigg(
- \frac{88}{9}  \z2
+ \bigg(
- \frac{484}{27}\bigg)  \Lqr
+ \frac{3560}{81}\bigg)  \CA^2  \CF
+ \bigg(
- \frac{64}{9}  \z2
+ \bigg(
- \frac{136}{9}\bigg)  \Lfr
+ \bigg(\frac{8}{3}\bigg)  \Lfr^2
+ \bigg(\frac{16}{3}\bigg)  \Lqrfr
+ \bigg(\frac{160}{9}\bigg)  \Lqr
+ \bigg(-8\bigg)  \Lqr^2
- \frac{212}{27}\bigg)  \CF^2  \NF
+ \bigg(\frac{16}{9}  \z2
+ \bigg(\frac{176}{27}\bigg)  \Lqr
- \frac{1156}{81}\bigg)   \CA  \CF  \NF
+ \bigg(\frac{352}{9}  \z2
- 56  \z3
+ \bigg(
- 32  \z2
+ \frac{940}{9}\bigg)  \Lfr
+ \bigg(32  \z2
- \frac{1072}{9}\bigg)  \Lqr
+ \bigg(
- \frac{88}{3}\bigg)  \Lqrfr
+ \bigg(
- \frac{44}{3}\bigg)  \Lfr^2
+ \bigg(44\bigg)  \Lqr^2
+ \frac{1352}{27}\bigg)  \CA   \CF^2
+ \bigg(\bigg(
- 128  \z2
+ 32\bigg)  \Lqr
+ \bigg(128  \z2
- 32\bigg)  \Lfr
+ \bigg(\bigg(-32\bigg)  \Lfr
+ \bigg(32\bigg)   \Lqr
+ 48\bigg)  \Lqrfr
+ \bigg(
- \frac{32}{3}\bigg)  \Lqr^3
+ \bigg(\frac{32}{3}\bigg)  \Lfr^3
+ \bigg(-48\bigg)  \Lfr^2\bigg)  \CF^3
+ \bigg(\bigg(
- \frac{16}{27}\bigg)  \Lqr
+ \frac{80}{81}\bigg)  \CF  \NF^2 \bigg\}      
+ \LNoNtb^2    \bigg\{ \bigg(
- \frac{504}{5}  \z2^2
+ \frac{3824}{9}  \z2
+ \frac{560}{9}  \z3
+ \bigg(
- 176  \z2
+ 64   \z3
- \frac{3088}{27}\bigg)  \Lqr
+ \bigg(
- \frac{104}{3}  \z2
- 64  \z3
+ \frac{514}{27}\bigg)  \Lfr
+ \bigg(
- 32  \z2
+ \frac{1072}{9}\bigg)  \Lqr^2
+ \bigg(
- 32  \z2
+ \frac{1270}{9}\bigg)  \Lfr^2
+ \bigg(64  \z2
+ \bigg(\frac{88}{3}\bigg)  \Lqr
+ \bigg(\frac{88}{3}\bigg)  \Lfr
- \frac{1748}{9}\bigg)  \Lqrfr
+ \bigg(
- \frac{88}{3}\bigg)  \Lqr^3
+ \bigg(
- \frac{88}{3}\bigg)  \Lfr^3
- \frac{68}{3}\bigg)  \CA   \CF^2
+ \bigg(\frac{88}{5}  \z2^2
- \frac{536}{9}  \z2
- 88  \z3
+ \bigg(\frac{88}{3}  \z2
- \frac{3560}{27}\bigg)  \Lqr
+ \bigg( \frac{242}{9}\bigg)  \Lqr^2
+ \frac{15503}{81}\bigg)  \CA^2  \CF
+ \bigg(\frac{1104}{5}  \z2^2
- 64  \z2
- 120  \z3
+ \bigg(
- 256  \z2
+ \bigg(-48\bigg)  \Lqr
+ \bigg(96\bigg)  \Lfr
+ 64\bigg)  \Lqrfr
+ \bigg(
- 144  \z2
- 96  \z3
+ 42\bigg)  \Lfr
+ \bigg(
- 48  \z2
+ 96  \z3\bigg)  \Lqr
+ \bigg(128  \z2
- 32\bigg)  \Lqr^2
+ \bigg(128  \z2
+ 4\bigg)   \Lfr^2
+ \bigg(-48\bigg)  \Lfr^3
+ 32\bigg)  \CF^3
+ \bigg(
- \frac{560}{9}  \z2
+ \frac{160}{9}  \z3
+ \bigg(\frac{32}{3}  \z2
- \frac{52}{27}\bigg)  \Lfr
+ \bigg(32  \z2
+ \frac{412}{27}\bigg)  \Lqr
+ \bigg(\bigg(
- \frac{16}{3}\bigg)  \Lqr
+ \bigg(
- \frac{16}{3}\bigg)  \Lfr
+ \frac{248}{9}\bigg)  \Lqrfr
+ \bigg(
- \frac{196}{9}\bigg)  \Lfr^2
+ \bigg(
- \frac{160}{9}\bigg)  \Lqr^2
+ \bigg(\frac{16}{3}\bigg)  \Lqr^3
+ \bigg(\frac{16}{3}\bigg)  \Lfr^3
- \frac{23}{3}\bigg)  \CF^2  \NF
+ \bigg(\frac{80}{9}  \z2
+ \bigg(
- \frac{16}{3}  \z2
+ \frac{1156}{27}\bigg)   \Lqr
+ \bigg(
- \frac{88}{9}\bigg)  \Lqr^2
- \frac{4102}{81}\bigg)  \CA  \CF  \NF
+ \bigg(\bigg(
- \frac{80}{27}\bigg)  \Lqr
+ \bigg(\frac{8}{9}\bigg)   \Lqr^2
+ \frac{200}{81}\bigg)  \CF  \NF^2 \bigg\}      
+ \LNoNtb    \bigg\{ \bigg(
- \frac{88}{5}  \z2^2
+ \frac{176}{3}  \z2  \z3
- \frac{6392}{81}  \z2
- \frac{12328}{27}  \z3
+ 192  \z5
+ \bigg(
- \frac{176}{5}  \z2^2
+ \frac{1072}{9}  \z2
+ 176  \z3
- \frac{31006}{81}\bigg)  \Lqr
+ \bigg(\frac{176}{5}   \z2^2
- \frac{1072}{9}  \z2
+ \frac{88}{3}  \z3
+ \frac{490}{3}\bigg)  \Lfr
+ \bigg(
- \frac{88}{3}  \z2
+ \frac{3560}{27}\bigg)   \Lqr^2
+ \bigg(\frac{88}{3}  \z2
- \frac{3560}{27}\bigg)  \Lfr^2
+ \bigg(
- \frac{484}{27}\bigg)  \Lqr^3
+ \bigg(\frac{484}{27}\bigg)   \Lfr^3
+ \frac{297029}{729}\bigg)  \CA^2  \CF
+ \bigg(
- \frac{16}{5}  \z2^2
+ \frac{824}{81}  \z2
+ \frac{904}{27}  \z3
+ \bigg(
- \frac{160}{9}  \z2
+ \frac{8204}{81}\bigg)  \Lqr
+ \bigg(
- \frac{16}{3}  \z2
+ \frac{1156}{27}\bigg)  \Lfr^2
+ \bigg(\frac{16}{3}   \z2
- \frac{1156}{27}\bigg)  \Lqr^2
+ \bigg(\frac{160}{9}  \z2
- \frac{112}{3}  \z3
- \frac{836}{27}\bigg)  \Lfr
+ \bigg(
- \frac{176}{27}\bigg)  \Lfr^3
+ \bigg(\frac{176}{27}\bigg)  \Lqr^3
- \frac{62626}{729}\bigg)  \CA  \CF  \NF
+ \bigg(\frac{32}{5}  \z2^2
- \frac{1792}{27}  \z2
+ \frac{304}{9}  \z3
+ \bigg(
- \frac{1120}{9}  \z2
+ \frac{320}{9}  \z3
+ \frac{86}{9}\bigg)  \Lfr
+ \bigg(
- \frac{128}{3}   \z2
+ \frac{4}{3}\bigg)  \Lqr^2
+ \bigg(\frac{128}{3}  \z2
+ 28\bigg)  \Lfr^2
+ \bigg(\frac{1120}{9}  \z2
- \frac{320}{9}  \z3
+ \frac{46}{3}\bigg)  \Lqr
+ \bigg(\bigg(8\bigg)  \Lqr
+ \bigg(8\bigg)  \Lfr
- \frac{88}{3}\bigg)  \Lqrfr
+ \bigg(-16\bigg)  \Lfr^3
- \frac{421}{9}\bigg)  \CF^2   \NF
+ \bigg(
- 448  \z2  \z3
+ \frac{12928}{27}  \z2
+ 112  \z3
+ \bigg(
- \frac{1008}{5}  \z2^2
+ \frac{7648}{9}   \z2
+ \frac{2632}{9}  \z3
- \frac{2024}{9}\bigg)  \Lfr
+ \bigg(\frac{1008}{5}  \z2^2
- \frac{7648}{9}  \z2
- \frac{1120}{9}  \z3
+ \frac{136}{3}\bigg)  \Lqr
+ \bigg(
- \frac{560}{3}  \z2
+ 96  \z3
- 172\bigg)  \Lfr^2
+ \bigg(
- 48  \z2
- 192   \z3
+ \bigg(-44\bigg)  \Lqr
+ \bigg(-44\bigg)  \Lfr
+ \frac{460}{3}\bigg)  \Lqrfr
+ \bigg(\frac{704}{3}  \z2
+ 96  \z3
+ \frac{56}{3}\bigg)   \Lqr^2
+ \bigg(88\bigg)  \Lfr^3
- \frac{3232}{27}\bigg)  \CA  \CF^2
+ \bigg(\frac{32}{9}  \z3
+ \bigg(
- \frac{400}{81}\bigg)  \Lqr
+ \bigg(
- \frac{80}{27}\bigg)  \Lfr^2
+ \bigg(
- \frac{16}{27}\bigg)  \Lqr^3
+ \bigg(
- \frac{16}{27}\bigg)  \Lfr
+ \bigg(\frac{16}{27}\bigg)   \Lfr^3
+ \bigg(\frac{80}{27}\bigg)  \Lqr^2
+ \frac{1856}{729}\bigg)  \CF  \NF^2
+ \bigg(\bigg(
- \frac{2208}{5}  \z2^2
+ 128  \z2
+ 240  \z3
- 64\bigg)  \Lqr
+ \bigg(\frac{2208}{5}  \z2^2
- 128  \z2
- 240  \z3
+ 64\bigg)  \Lfr
+ \bigg(
- 288  \z2
- 192  \z3
+ 84\bigg)  \Lfr^2
+ \bigg(96  \z2
- 192  \z3\bigg)  \Lqr^2
+ \bigg(192  \z2
+ 384   \z3
+ \bigg(-72\bigg)  \Lfr
- 84\bigg)  \Lqrfr
+ \bigg(72\bigg)  \Lfr^3\bigg)  \CF^3 \bigg\} 
+ \g03bbH\,.
\end{autobreak} 
\end{align}

\bibliographystyle{JHEP}
\bibliography{bbHrapres} 

\providecommand{\href}[2]{#2}\begingroup\raggedright\begin{thebibliography}{100}

\bibitem{CMS:2018nsn}
{\scshape CMS} collaboration, A.~M. Sirunyan et~al., \emph{{Observation of
  Higgs boson decay to bottom quarks}},
  \href{https://doi.org/10.1103/PhysRevLett.121.121801}{\emph{Phys. Rev. Lett.}
  {\bfseries 121} (2018) 121801}
  [\href{https://arxiv.org/abs/1808.08242}{{\ttfamily 1808.08242}}].

\bibitem{ATLAS:2018kot}
{\scshape ATLAS} collaboration, M.~Aaboud et~al., \emph{{Observation of $H
  \rightarrow b\bar{b}$ decays and $VH$ production with the ATLAS detector}},
  \href{https://doi.org/10.1016/j.physletb.2018.09.013}{\emph{Phys. Lett. B}
  {\bfseries 786} (2018) 59}
  [\href{https://arxiv.org/abs/1808.08238}{{\ttfamily 1808.08238}}].

\bibitem{ATLAS:2018hxb}
{\scshape ATLAS} collaboration, M.~Aaboud et~al., \emph{{Measurements of Higgs
  boson properties in the diphoton decay channel with 36 fb$^{-1}$ of $pp$
  collision data at $\sqrt{s} = 13$ TeV with the ATLAS detector}},
  \href{https://doi.org/10.1103/PhysRevD.98.052005}{\emph{Phys. Rev. D}
  {\bfseries 98} (2018) 052005}
  [\href{https://arxiv.org/abs/1802.04146}{{\ttfamily 1802.04146}}].

\bibitem{CMS:2020xrn}
{\scshape CMS} collaboration, A.~M. Sirunyan et~al., \emph{{A measurement of
  the Higgs boson mass in the diphoton decay channel}},
  \href{https://doi.org/10.1016/j.physletb.2020.135425}{\emph{Phys. Lett. B}
  {\bfseries 805} (2020) 135425}
  [\href{https://arxiv.org/abs/2002.06398}{{\ttfamily 2002.06398}}].

\bibitem{ATLAS:2019qet}
{\scshape ATLAS} collaboration, M.~Aaboud et~al., \emph{{Measurement of the
  four-lepton invariant mass spectrum in 13 TeV proton-proton collisions with
  the ATLAS detector}},
  \href{https://doi.org/10.1007/JHEP04(2019)048}{\emph{JHEP} {\bfseries 04}
  (2019) 048} [\href{https://arxiv.org/abs/1902.05892}{{\ttfamily
  1902.05892}}].

\bibitem{CMS:2021ugl}
{\scshape CMS} collaboration, A.~M. Sirunyan et~al., \emph{{Measurements of
  production cross sections of the Higgs boson in the four-lepton final state
  in proton\textendash{}proton collisions at $\sqrt{s} = 13\,\text {Te}\text
  {V} $}}, \href{https://doi.org/10.1140/epjc/s10052-021-09200-x}{\emph{Eur.
  Phys. J. C} {\bfseries 81} (2021) 488}
  [\href{https://arxiv.org/abs/2103.04956}{{\ttfamily 2103.04956}}].

\bibitem{Anastasiou:2015vya}
C.~Anastasiou, C.~Duhr, F.~Dulat, F.~Herzog and B.~Mistlberger, \emph{{Higgs
  Boson Gluon-Fusion Production in QCD at Three Loops}},
  \href{https://doi.org/10.1103/PhysRevLett.114.212001}{\emph{Phys. Rev. Lett.}
  {\bfseries 114} (2015) 212001}
  [\href{https://arxiv.org/abs/1503.06056}{{\ttfamily 1503.06056}}].

\bibitem{Mistlberger:2018etf}
B.~Mistlberger, \emph{{Higgs boson production at hadron colliders at N$^{3}$LO
  in QCD}}, \href{https://doi.org/10.1007/JHEP05(2018)028}{\emph{JHEP}
  {\bfseries 05} (2018) 028}
  [\href{https://arxiv.org/abs/1802.00833}{{\ttfamily 1802.00833}}].

\bibitem{Harlander:2002wh}
R.~V. Harlander and W.~B. Kilgore, \emph{{Next-to-next-to-leading order Higgs
  production at hadron colliders}},
  \href{https://doi.org/10.1103/PhysRevLett.88.201801}{\emph{Phys. Rev. Lett.}
  {\bfseries 88} (2002) 201801}
  [\href{https://arxiv.org/abs/hep-ph/0201206}{{\ttfamily hep-ph/0201206}}].

\bibitem{Anastasiou:2002yz}
C.~Anastasiou and K.~Melnikov, \emph{{Higgs boson production at hadron
  colliders in NNLO QCD}},
  \href{https://doi.org/10.1016/S0550-3213(02)00837-4}{\emph{Nucl. Phys. B}
  {\bfseries 646} (2002) 220}
  [\href{https://arxiv.org/abs/hep-ph/0207004}{{\ttfamily hep-ph/0207004}}].

\bibitem{Ravindran:2003um}
V.~Ravindran, J.~Smith and W.~L. van Neerven, \emph{{NNLO corrections to the
  total cross-section for Higgs boson production in hadron hadron collisions}},
  \href{https://doi.org/10.1016/S0550-3213(03)00457-7}{\emph{Nucl. Phys. B}
  {\bfseries 665} (2003) 325}
  [\href{https://arxiv.org/abs/hep-ph/0302135}{{\ttfamily hep-ph/0302135}}].

\bibitem{Bolzoni:2010xr}
P.~Bolzoni, F.~Maltoni, S.~Moch and M.~Zaro, \emph{{Higgs production via
  vector-boson fusion at NNLO in QCD}},
  \href{https://doi.org/10.1103/PhysRevLett.105.011801}{\emph{Phys. Rev. Lett.}
  {\bfseries 105} (2010) 011801}
  [\href{https://arxiv.org/abs/1003.4451}{{\ttfamily 1003.4451}}].

\bibitem{Bolzoni:2011cu}
P.~Bolzoni, F.~Maltoni, S.~Moch and M.~Zaro, \emph{{Vector boson fusion at NNLO
  in QCD: SM Higgs and beyond}},
  \href{https://doi.org/10.1103/PhysRevD.85.035002}{\emph{Phys. Rev. D}
  {\bfseries 85} (2012) 035002}
  [\href{https://arxiv.org/abs/1109.3717}{{\ttfamily 1109.3717}}].

\bibitem{Dreyer:2016oyx}
F.~A. Dreyer and A.~Karlberg, \emph{{Vector-Boson Fusion Higgs Production at
  Three Loops in QCD}},
  \href{https://doi.org/10.1103/PhysRevLett.117.072001}{\emph{Phys. Rev. Lett.}
  {\bfseries 117} (2016) 072001}
  [\href{https://arxiv.org/abs/1606.00840}{{\ttfamily 1606.00840}}].

\bibitem{Buckley:2021gfw}
A.~Buckley et~al., \emph{{A comparative study of Higgs boson production from
  vector-boson fusion}},
  \href{https://doi.org/10.1007/JHEP11(2021)108}{\emph{JHEP} {\bfseries 11}
  (2021) 108} [\href{https://arxiv.org/abs/2105.11399}{{\ttfamily
  2105.11399}}].

\bibitem{Das:2019btv}
G.~Das, S.~Moch and A.~Vogt, \emph{{Soft corrections to inclusive
  deep-inelastic scattering at four loops and beyond}},
  \href{https://doi.org/10.1007/JHEP03(2020)116}{\emph{JHEP} {\bfseries 03}
  (2020) 116} [\href{https://arxiv.org/abs/1912.12920}{{\ttfamily
  1912.12920}}].

\bibitem{Das:2020adl}
G.~Das, S.~Moch and A.~Vogt, \emph{{Approximate four-loop QCD corrections to
  the Higgs-boson production cross section}},
  \href{https://doi.org/10.1016/j.physletb.2020.135546}{\emph{Phys. Lett. B}
  {\bfseries 807} (2020) 135546}
  [\href{https://arxiv.org/abs/2004.00563}{{\ttfamily 2004.00563}}].

\bibitem{Catani:2003zt}
S.~Catani, D.~de~Florian, M.~Grazzini and P.~Nason, \emph{{Soft gluon
  resummation for Higgs boson production at hadron colliders}},
  \href{https://doi.org/10.1088/1126-6708/2003/07/028}{\emph{JHEP} {\bfseries
  07} (2003) 028} [\href{https://arxiv.org/abs/hep-ph/0306211}{{\ttfamily
  hep-ph/0306211}}].

\bibitem{Moch:2005ky}
S.~Moch and A.~Vogt, \emph{{Higher-order soft corrections to lepton pair and
  Higgs boson production}},
  \href{https://doi.org/10.1016/j.physletb.2005.09.061}{\emph{Phys. Lett. B}
  {\bfseries 631} (2005) 48}
  [\href{https://arxiv.org/abs/hep-ph/0508265}{{\ttfamily hep-ph/0508265}}].

\bibitem{Laenen:2005uz}
E.~Laenen and L.~Magnea, \emph{{Threshold resummation for electroweak
  annihilation from DIS data}},
  \href{https://doi.org/10.1016/j.physletb.2005.10.038}{\emph{Phys. Lett. B}
  {\bfseries 632} (2006) 270}
  [\href{https://arxiv.org/abs/hep-ph/0508284}{{\ttfamily hep-ph/0508284}}].

\bibitem{Idilbi:2005ni}
A.~Idilbi, X.-d. Ji, J.-P. Ma and F.~Yuan, \emph{{Threshold resummation for
  Higgs production in effective field theory}},
  \href{https://doi.org/10.1103/PhysRevD.73.077501}{\emph{Phys. Rev. D}
  {\bfseries 73} (2006) 077501}
  [\href{https://arxiv.org/abs/hep-ph/0509294}{{\ttfamily hep-ph/0509294}}].

\bibitem{Bonvini:2014joa}
M.~Bonvini and S.~Marzani, \emph{{Resummed Higgs cross section at N$^{3}$LL}},
  \href{https://doi.org/10.1007/JHEP09(2014)007}{\emph{JHEP} {\bfseries 09}
  (2014) 007} [\href{https://arxiv.org/abs/1405.3654}{{\ttfamily 1405.3654}}].

\bibitem{Bonvini:2016frm}
M.~Bonvini, S.~Marzani, C.~Muselli and L.~Rottoli, \emph{{On the Higgs cross
  section at N$^{3}$LO+N$^{3}$LL and its uncertainty}},
  \href{https://doi.org/10.1007/JHEP08(2016)105}{\emph{JHEP} {\bfseries 08}
  (2016) 105} [\href{https://arxiv.org/abs/1603.08000}{{\ttfamily
  1603.08000}}].

\bibitem{Ebert:2017uel}
M.~A. Ebert, J.~K.~L. Michel and F.~J. Tackmann, \emph{{Resummation Improved
  Rapidity Spectrum for Gluon Fusion Higgs Production}},
  \href{https://doi.org/10.1007/JHEP05(2017)088}{\emph{JHEP} {\bfseries 05}
  (2017) 088} [\href{https://arxiv.org/abs/1702.00794}{{\ttfamily
  1702.00794}}].

\bibitem{Graudenz:1992pv}
D.~Graudenz, M.~Spira and P.~M. Zerwas, \emph{{QCD corrections to Higgs boson
  production at proton proton colliders}},
  \href{https://doi.org/10.1103/PhysRevLett.70.1372}{\emph{Phys. Rev. Lett.}
  {\bfseries 70} (1993) 1372}.

\bibitem{Spira:1995rr}
M.~Spira, A.~Djouadi, D.~Graudenz and P.~M. Zerwas, \emph{{Higgs boson
  production at the LHC}},
  \href{https://doi.org/10.1016/0550-3213(95)00379-7}{\emph{Nucl. Phys. B}
  {\bfseries 453} (1995) 17}
  [\href{https://arxiv.org/abs/hep-ph/9504378}{{\ttfamily hep-ph/9504378}}].

\bibitem{Marzani:2008az}
S.~Marzani, R.~D. Ball, V.~Del~Duca, S.~Forte and A.~Vicini, \emph{{Higgs
  production via gluon-gluon fusion with finite top mass beyond next-to-leading
  order}}, \href{https://doi.org/10.1016/j.nuclphysb.2008.03.016}{\emph{Nucl.
  Phys. B} {\bfseries 800} (2008) 127}
  [\href{https://arxiv.org/abs/0801.2544}{{\ttfamily 0801.2544}}].

\bibitem{Pak:2009dg}
A.~Pak, M.~Rogal and M.~Steinhauser, \emph{{Finite top quark mass effects in
  NNLO Higgs boson production at LHC}},
  \href{https://doi.org/10.1007/JHEP02(2010)025}{\emph{JHEP} {\bfseries 02}
  (2010) 025} [\href{https://arxiv.org/abs/0911.4662}{{\ttfamily 0911.4662}}].

\bibitem{Harlander:2009my}
R.~V. Harlander, H.~Mantler, S.~Marzani and K.~J. Ozeren, \emph{{Higgs
  production in gluon fusion at next-to-next-to-leading order QCD for finite
  top mass}}, \href{https://doi.org/10.1140/epjc/s10052-010-1258-x}{\emph{Eur.
  Phys. J. C} {\bfseries 66} (2010) 359}
  [\href{https://arxiv.org/abs/0912.2104}{{\ttfamily 0912.2104}}].

\bibitem{Harlander:2009mq}
R.~V. Harlander and K.~J. Ozeren, \emph{{Finite top mass effects for hadronic
  Higgs production at next-to-next-to-leading order}},
  \href{https://doi.org/10.1088/1126-6708/2009/11/088}{\emph{JHEP} {\bfseries
  11} (2009) 088} [\href{https://arxiv.org/abs/0909.3420}{{\ttfamily
  0909.3420}}].

\bibitem{Czakon:2021yub}
M.~Czakon, R.~V. Harlander, J.~Klappert and M.~Niggetiedt, \emph{{Exact
  Top-Quark Mass Dependence in Hadronic Higgs Production}},
  \href{https://doi.org/10.1103/PhysRevLett.127.162002}{\emph{Phys. Rev. Lett.}
  {\bfseries 127} (2021) 162002}
  [\href{https://arxiv.org/abs/2105.04436}{{\ttfamily 2105.04436}}].

\bibitem{Aglietti:2004nj}
U.~Aglietti, R.~Bonciani, G.~Degrassi and A.~Vicini, \emph{{Two loop light
  fermion contribution to Higgs production and decays}},
  \href{https://doi.org/10.1016/j.physletb.2004.06.063}{\emph{Phys. Lett. B}
  {\bfseries 595} (2004) 432}
  [\href{https://arxiv.org/abs/hep-ph/0404071}{{\ttfamily hep-ph/0404071}}].

\bibitem{Actis:2008ug}
S.~Actis, G.~Passarino, C.~Sturm and S.~Uccirati, \emph{{NLO Electroweak
  Corrections to Higgs Boson Production at Hadron Colliders}},
  \href{https://doi.org/10.1016/j.physletb.2008.10.018}{\emph{Phys. Lett. B}
  {\bfseries 670} (2008) 12} [\href{https://arxiv.org/abs/0809.1301}{{\ttfamily
  0809.1301}}].

\bibitem{Gehrmann:2014vha}
T.~Gehrmann and D.~Kara, \emph{{The $Hb\bar{b}$ form factor to three loops in
  QCD}}, \href{https://doi.org/10.1007/JHEP09(2014)174}{\emph{JHEP} {\bfseries
  09} (2014) 174} [\href{https://arxiv.org/abs/1407.8114}{{\ttfamily
  1407.8114}}].

\bibitem{Anastasiou:2014vaa}
C.~Anastasiou, C.~Duhr, F.~Dulat, E.~Furlan, T.~Gehrmann, F.~Herzog et~al.,
  \emph{{Higgs boson gluon\textendash{}fusion production at threshold in
  N$^3$LO QCD}},
  \href{https://doi.org/10.1016/j.physletb.2014.08.067}{\emph{Phys. Lett. B}
  {\bfseries 737} (2014) 325}
  [\href{https://arxiv.org/abs/1403.4616}{{\ttfamily 1403.4616}}].

\bibitem{Vogt:2004mw}
A.~Vogt, S.~Moch and J.~A.~M. Vermaseren, \emph{{The Three-loop splitting
  functions in QCD: The Singlet case}},
  \href{https://doi.org/10.1016/j.nuclphysb.2004.04.024}{\emph{Nucl. Phys. B}
  {\bfseries 691} (2004) 129}
  [\href{https://arxiv.org/abs/hep-ph/0404111}{{\ttfamily hep-ph/0404111}}].

\bibitem{Moch:2004pa}
S.~Moch, J.~A.~M. Vermaseren and A.~Vogt, \emph{{The Three loop splitting
  functions in QCD: The Nonsinglet case}},
  \href{https://doi.org/10.1016/j.nuclphysb.2004.03.030}{\emph{Nucl. Phys. B}
  {\bfseries 688} (2004) 101}
  [\href{https://arxiv.org/abs/hep-ph/0403192}{{\ttfamily hep-ph/0403192}}].

\bibitem{Dicus:1998hs}
D.~Dicus, T.~Stelzer, Z.~Sullivan and S.~Willenbrock, \emph{{Higgs boson
  production in association with bottom quarks at next-to-leading order}},
  \href{https://doi.org/10.1103/PhysRevD.59.094016}{\emph{Phys. Rev. D}
  {\bfseries 59} (1999) 094016}
  [\href{https://arxiv.org/abs/hep-ph/9811492}{{\ttfamily hep-ph/9811492}}].

\bibitem{Balazs:1998sb}
C.~Balazs, H.-J. He and C.~P. Yuan, \emph{{QCD corrections to scalar production
  via heavy quark fusion at hadron colliders}},
  \href{https://doi.org/10.1103/PhysRevD.60.114001}{\emph{Phys. Rev. D}
  {\bfseries 60} (1999) 114001}
  [\href{https://arxiv.org/abs/hep-ph/9812263}{{\ttfamily hep-ph/9812263}}].

\bibitem{Maltoni:2003pn}
F.~Maltoni, Z.~Sullivan and S.~Willenbrock, \emph{{Higgs-Boson Production via
  Bottom-Quark Fusion}},
  \href{https://doi.org/10.1103/PhysRevD.67.093005}{\emph{Phys. Rev. D}
  {\bfseries 67} (2003) 093005}
  [\href{https://arxiv.org/abs/hep-ph/0301033}{{\ttfamily hep-ph/0301033}}].

\bibitem{Harlander:2003ai}
R.~V. Harlander and W.~B. Kilgore, \emph{{Higgs boson production in bottom
  quark fusion at next-to-next-to leading order}},
  \href{https://doi.org/10.1103/PhysRevD.68.013001}{\emph{Phys. Rev. D}
  {\bfseries 68} (2003) 013001}
  [\href{https://arxiv.org/abs/hep-ph/0304035}{{\ttfamily hep-ph/0304035}}].

\bibitem{Duhr:2019kwi}
C.~Duhr, F.~Dulat and B.~Mistlberger, \emph{{Higgs Boson Production in
  Bottom-Quark Fusion to Third Order in the Strong Coupling}},
  \href{https://doi.org/10.1103/PhysRevLett.125.051804}{\emph{Phys. Rev. Lett.}
  {\bfseries 125} (2020) 051804}
  [\href{https://arxiv.org/abs/1904.09990}{{\ttfamily 1904.09990}}].

\bibitem{Dittmaier:2003ej}
S.~Dittmaier, M.~Kr\"amer and M.~Spira, \emph{{Higgs radiation off bottom
  quarks at the Tevatron and the CERN LHC}},
  \href{https://doi.org/10.1103/PhysRevD.70.074010}{\emph{Phys. Rev. D}
  {\bfseries 70} (2004) 074010}
  [\href{https://arxiv.org/abs/hep-ph/0309204}{{\ttfamily hep-ph/0309204}}].

\bibitem{Dawson:2003kb}
S.~Dawson, C.~B. Jackson, L.~Reina and D.~Wackeroth, \emph{{Exclusive Higgs
  boson production with bottom quarks at hadron colliders}},
  \href{https://doi.org/10.1103/PhysRevD.69.074027}{\emph{Phys. Rev. D}
  {\bfseries 69} (2004) 074027}
  [\href{https://arxiv.org/abs/hep-ph/0311067}{{\ttfamily hep-ph/0311067}}].

\bibitem{Wiesemann:2014ioa}
M.~Wiesemann, R.~Frederix, S.~Frixione, V.~Hirschi, F.~Maltoni and
  P.~Torrielli, \emph{{Higgs production in association with bottom quarks}},
  \href{https://doi.org/10.1007/JHEP02(2015)132}{\emph{JHEP} {\bfseries 02}
  (2015) 132} [\href{https://arxiv.org/abs/1409.5301}{{\ttfamily 1409.5301}}].

\bibitem{Aivazis:1993pi}
M.~A.~G. Aivazis, J.~C. Collins, F.~I. Olness and W.-K. Tung,
  \emph{{Leptoproduction of heavy quarks. 2. A Unified QCD formulation of
  charged and neutral current processes from fixed target to collider
  energies}}, \href{https://doi.org/10.1103/PhysRevD.50.3102}{\emph{Phys. Rev.
  D} {\bfseries 50} (1994) 3102}
  [\href{https://arxiv.org/abs/hep-ph/9312319}{{\ttfamily hep-ph/9312319}}].

\bibitem{Cacciari:1998it}
M.~Cacciari, M.~Greco and P.~Nason, \emph{{The $P_T$ spectrum in heavy flavor
  hadroproduction}},
  \href{https://doi.org/10.1088/1126-6708/1998/05/007}{\emph{JHEP} {\bfseries
  05} (1998) 007} [\href{https://arxiv.org/abs/hep-ph/9803400}{{\ttfamily
  hep-ph/9803400}}].

\bibitem{Forte:2010ta}
S.~Forte, E.~Laenen, P.~Nason and J.~Rojo, \emph{{Heavy quarks in
  deep-inelastic scattering}},
  \href{https://doi.org/10.1016/j.nuclphysb.2010.03.014}{\emph{Nucl. Phys. B}
  {\bfseries 834} (2010) 116}
  [\href{https://arxiv.org/abs/1001.2312}{{\ttfamily 1001.2312}}].

\bibitem{Harlander:2011aa}
R.~Harlander, M.~Kramer and M.~Schumacher, \emph{{Bottom-quark associated
  Higgs-boson production: reconciling the four- and five-flavour scheme
  approach}},  \href{https://arxiv.org/abs/1112.3478}{{\ttfamily 1112.3478}}.

\bibitem{Bonvini:2015pxa}
M.~Bonvini, A.~S. Papanastasiou and F.~J. Tackmann, \emph{{Resummation and
  matching of b-quark mass effects in $ b\overline{b}H $ production}},
  \href{https://doi.org/10.1007/JHEP11(2015)196}{\emph{JHEP} {\bfseries 11}
  (2015) 196} [\href{https://arxiv.org/abs/1508.03288}{{\ttfamily
  1508.03288}}].

\bibitem{Forte:2015hba}
S.~Forte, D.~Napoletano and M.~Ubiali, \emph{{Higgs production in bottom-quark
  fusion in a matched scheme}},
  \href{https://doi.org/10.1016/j.physletb.2015.10.051}{\emph{Phys. Lett. B}
  {\bfseries 751} (2015) 331}
  [\href{https://arxiv.org/abs/1508.01529}{{\ttfamily 1508.01529}}].

\bibitem{Bonvini:2016fgf}
M.~Bonvini, A.~S. Papanastasiou and F.~J. Tackmann, \emph{{Matched predictions
  for the $ b\overline{b}H $ cross section at the 13 TeV LHC}},
  \href{https://doi.org/10.1007/JHEP10(2016)053}{\emph{JHEP} {\bfseries 10}
  (2016) 053} [\href{https://arxiv.org/abs/1605.01733}{{\ttfamily
  1605.01733}}].

\bibitem{Forte:2016sja}
S.~Forte, D.~Napoletano and M.~Ubiali, \emph{{Higgs production in bottom-quark
  fusion: matching beyond leading order}},
  \href{https://doi.org/10.1016/j.physletb.2016.10.040}{\emph{Phys. Lett. B}
  {\bfseries 763} (2016) 190}
  [\href{https://arxiv.org/abs/1607.00389}{{\ttfamily 1607.00389}}].

\bibitem{Duhr:2020kzd}
C.~Duhr, F.~Dulat, V.~Hirschi and B.~Mistlberger, \emph{{Higgs production in
  bottom quark fusion: matching the 4- and 5-flavour schemes to third order in
  the strong coupling}},
  \href{https://doi.org/10.1007/JHEP08(2020)017}{\emph{JHEP} {\bfseries 08}
  (2020) 017} [\href{https://arxiv.org/abs/2004.04752}{{\ttfamily
  2004.04752}}].

\bibitem{Ajjath:2019neu}
A.~H. Ajjath, A.~Chakraborty, G.~Das, P.~Mukherjee and V.~Ravindran,
  \emph{{Resummed prediction for Higgs boson production through $b \overline{b}
  $ annihilation at N$^{3}$LL}},
  \href{https://doi.org/10.1007/JHEP11(2019)006}{\emph{JHEP} {\bfseries 11}
  (2019) 006} [\href{https://arxiv.org/abs/1905.03771}{{\ttfamily
  1905.03771}}].

\bibitem{Das:2022zie}
G.~Das, C.~Dey, M.~C. Kumar and K.~Samanta, \emph{{Threshold enhanced cross
  sections for colorless productions}},
  \href{https://doi.org/10.1103/PhysRevD.107.034038}{\emph{Phys. Rev. D}
  {\bfseries 107} (2023) 034038}
  [\href{https://arxiv.org/abs/2210.17534}{{\ttfamily 2210.17534}}].

\bibitem{Ajjath:2019ixh}
A.~H. Ajjath, P.~Banerjee, A.~Chakraborty, P.~K. Dhani, P.~Mukherjee, N.~Rana
  et~al., \emph{{NNLO QCD$\oplus$QED corrections to Higgs production in bottom
  quark annihilation}},
  \href{https://doi.org/10.1103/PhysRevD.100.114016}{\emph{Phys. Rev. D}
  {\bfseries 100} (2019) 114016}
  [\href{https://arxiv.org/abs/1906.09028}{{\ttfamily 1906.09028}}].

\bibitem{Anastasiou:2004xq}
C.~Anastasiou, K.~Melnikov and F.~Petriello, \emph{{Higgs boson production at
  hadron colliders: Differential cross sections through next-to-next-to-leading
  order}}, \href{https://doi.org/10.1103/PhysRevLett.93.262002}{\emph{Phys.
  Rev. Lett.} {\bfseries 93} (2004) 262002}
  [\href{https://arxiv.org/abs/hep-ph/0409088}{{\ttfamily hep-ph/0409088}}].

\bibitem{Anastasiou:2005qj}
C.~Anastasiou, K.~Melnikov and F.~Petriello, \emph{{Fully differential Higgs
  boson production and the di-photon signal through next-to-next-to-leading
  order}}, \href{https://doi.org/10.1016/j.nuclphysb.2005.06.036}{\emph{Nucl.
  Phys. B} {\bfseries 724} (2005) 197}
  [\href{https://arxiv.org/abs/hep-ph/0501130}{{\ttfamily hep-ph/0501130}}].

\bibitem{Buhler:2012ytl}
S.~B\"uhler, F.~Herzog, A.~Lazopoulos and R.~M\"uller, \emph{{The fully
  differential hadronic production of a Higgs boson via bottom quark fusion at
  NNLO}}, \href{https://doi.org/10.1007/JHEP07(2012)115}{\emph{JHEP} {\bfseries
  07} (2012) 115} [\href{https://arxiv.org/abs/1204.4415}{{\ttfamily
  1204.4415}}].

\bibitem{Mondini:2021nck}
R.~Mondini and C.~Williams, \emph{{Bottom-induced contributions to Higgs plus
  jet at next-to-next-to-leading order}},
  \href{https://doi.org/10.1007/JHEP05(2021)045}{\emph{JHEP} {\bfseries 05}
  (2021) 045} [\href{https://arxiv.org/abs/2102.05487}{{\ttfamily
  2102.05487}}].

\bibitem{Ravindran:2006bu}
V.~Ravindran, J.~Smith and W.~L. van Neerven, \emph{{QCD threshold corrections
  to di-lepton and Higgs rapidity distributions beyond $N^{2}$ LO}},
  \href{https://doi.org/10.1016/j.nuclphysb.2007.01.005}{\emph{Nucl. Phys. B}
  {\bfseries 767} (2007) 100}
  [\href{https://arxiv.org/abs/hep-ph/0608308}{{\ttfamily hep-ph/0608308}}].

\bibitem{Ahmed:2014uya}
T.~Ahmed, M.~K. Mandal, N.~Rana and V.~Ravindran, \emph{{Rapidity Distributions
  in Drell-Yan and Higgs Productions at Threshold to Third Order in QCD}},
  \href{https://doi.org/10.1103/PhysRevLett.113.212003}{\emph{Phys. Rev. Lett.}
  {\bfseries 113} (2014) 212003}
  [\href{https://arxiv.org/abs/1404.6504}{{\ttfamily 1404.6504}}].

\bibitem{Ahmed:2014era}
T.~Ahmed, M.~K. Mandal, N.~Rana and V.~Ravindran, \emph{{Higgs Rapidity
  Distribution in $b {\bar b}$ Annihilation at Threshold in N$^{3}$LO QCD}},
  \href{https://doi.org/10.1007/JHEP02(2015)131}{\emph{JHEP} {\bfseries 02}
  (2015) 131} [\href{https://arxiv.org/abs/1411.5301}{{\ttfamily 1411.5301}}].

\bibitem{Cieri:2018oms}
L.~Cieri, X.~Chen, T.~Gehrmann, E.~W.~N. Glover and A.~Huss, \emph{{Higgs boson
  production at the LHC using the $q_T$ subtraction formalism at N$^3$LO QCD}},
  \href{https://doi.org/10.1007/JHEP02(2019)096}{\emph{JHEP} {\bfseries 02}
  (2019) 096} [\href{https://arxiv.org/abs/1807.11501}{{\ttfamily
  1807.11501}}].

\bibitem{Chen:2021vtu}
X.~Chen, T.~Gehrmann, N.~Glover, A.~Huss, T.-Z. Yang and H.~X. Zhu,
  \emph{{Dilepton Rapidity Distribution in Drell-Yan Production to Third Order
  in QCD}}, \href{https://doi.org/10.1103/PhysRevLett.128.052001}{\emph{Phys.
  Rev. Lett.} {\bfseries 128} (2022) 052001}
  [\href{https://arxiv.org/abs/2107.09085}{{\ttfamily 2107.09085}}].

\bibitem{Catani:2007vq}
S.~Catani and M.~Grazzini, \emph{{An NNLO subtraction formalism in hadron
  collisions and its application to Higgs boson production at the LHC}},
  \href{https://doi.org/10.1103/PhysRevLett.98.222002}{\emph{Phys. Rev. Lett.}
  {\bfseries 98} (2007) 222002}
  [\href{https://arxiv.org/abs/hep-ph/0703012}{{\ttfamily hep-ph/0703012}}].

\bibitem{Catani:2009sm}
S.~Catani, L.~Cieri, G.~Ferrera, D.~de~Florian and M.~Grazzini, \emph{{Vector
  boson production at hadron colliders: a fully exclusive QCD calculation at
  NNLO}}, \href{https://doi.org/10.1103/PhysRevLett.103.082001}{\emph{Phys.
  Rev. Lett.} {\bfseries 103} (2009) 082001}
  [\href{https://arxiv.org/abs/0903.2120}{{\ttfamily 0903.2120}}].

\bibitem{Catani:2010en}
S.~Catani, G.~Ferrera and M.~Grazzini, \emph{{W Boson Production at Hadron
  Colliders: The Lepton Charge Asymmetry in NNLO QCD}},
  \href{https://doi.org/10.1007/JHEP05(2010)006}{\emph{JHEP} {\bfseries 05}
  (2010) 006} [\href{https://arxiv.org/abs/1002.3115}{{\ttfamily 1002.3115}}].

\bibitem{Catani:2011qz}
S.~Catani, L.~Cieri, D.~de~Florian, G.~Ferrera and M.~Grazzini, \emph{{Diphoton
  production at hadron colliders: a fully-differential QCD calculation at
  NNLO}}, \href{https://doi.org/10.1103/PhysRevLett.108.072001}{\emph{Phys.
  Rev. Lett.} {\bfseries 108} (2012) 072001}
  [\href{https://arxiv.org/abs/1110.2375}{{\ttfamily 1110.2375}}].

\bibitem{Banerjee:2017cfc}
P.~Banerjee, G.~Das, P.~K. Dhani and V.~Ravindran, \emph{{Threshold resummation
  of the rapidity distribution for Higgs production at NNLO+NNLL}},
  \href{https://doi.org/10.1103/PhysRevD.97.054024}{\emph{Phys. Rev. D}
  {\bfseries 97} (2018) 054024}
  [\href{https://arxiv.org/abs/1708.05706}{{\ttfamily 1708.05706}}].

\bibitem{Ajjath:2020lwb}
A.~H. Ajjath, P.~Mukherjee, V.~Ravindran, A.~Sankar and S.~Tiwari,
  \emph{{Next-to-soft corrections for Drell-Yan and Higgs boson rapidity
  distributions beyond N$^3$LO}},
  \href{https://doi.org/10.1103/PhysRevD.103.L111502}{\emph{Phys. Rev. D}
  {\bfseries 103} (2021) L111502}
  [\href{https://arxiv.org/abs/2010.00079}{{\ttfamily 2010.00079}}].

\bibitem{Catani:1989ne}
S.~Catani and L.~Trentadue, \emph{{Resummation of the QCD Perturbative Series
  for Hard Processes}},
  \href{https://doi.org/10.1016/0550-3213(89)90273-3}{\emph{Nucl. Phys. B}
  {\bfseries 327} (1989) 323}.

\bibitem{Westmark:2017uig}
D.~Westmark and J.~F. Owens, \emph{{Enhanced threshold resummation formalism
  for lepton pair production and its effects in the determination of parton
  distribution functions}},
  \href{https://doi.org/10.1103/PhysRevD.95.056024}{\emph{Phys. Rev. D}
  {\bfseries 95} (2017) 056024}
  [\href{https://arxiv.org/abs/1701.06716}{{\ttfamily 1701.06716}}].

\bibitem{Banerjee:2018vvb}
P.~Banerjee, G.~Das, P.~K. Dhani and V.~Ravindran, \emph{{Threshold resummation
  of the rapidity distribution for Drell-Yan production at NNLO+NNLL}},
  \href{https://doi.org/10.1103/PhysRevD.98.054018}{\emph{Phys. Rev. D}
  {\bfseries 98} (2018) 054018}
  [\href{https://arxiv.org/abs/1805.01186}{{\ttfamily 1805.01186}}].

\bibitem{Banerjee:2018mkm}
P.~Banerjee, G.~Das, P.~K. Dhani and V.~Ravindran, \emph{{Threshold resummation
  in the rapidity distribution for a colorless particle production at the
  LHC}}, \href{https://doi.org/10.22323/1.303.0043}{\emph{PoS} {\bfseries
  LL2018} (2018) 043} [\href{https://arxiv.org/abs/1807.04583}{{\ttfamily
  1807.04583}}].

\bibitem{Ahmed:2020amh}
T.~Ahmed, A.~A. H., P.~Mukherjee, V.~Ravindran and A.~Sankar, \emph{{Rapidity
  distribution at soft-virtual and beyond for $n$-colorless particles to
  ${N}^4$LO in QCD}},
  \href{https://doi.org/10.1140/epjc/s10052-021-09658-9}{\emph{Eur. Phys. J. C}
  {\bfseries 81} (2021) 943}
  [\href{https://arxiv.org/abs/2010.02980}{{\ttfamily 2010.02980}}].

\bibitem{Ravindran:2007sv}
V.~Ravindran and J.~Smith, \emph{{Threshold corrections to rapidity
  distributions of $Z$ and $W^\pm$ bosons beyond $N^{2}$LO at hadron
  colliders}}, \href{https://doi.org/10.1103/PhysRevD.76.114004}{\emph{Phys.
  Rev. D} {\bfseries 76} (2007) 114004}
  [\href{https://arxiv.org/abs/0708.1689}{{\ttfamily 0708.1689}}].

\bibitem{Lustermans:2019cau}
G.~Lustermans, J.~K.~L. Michel and F.~J. Tackmann, \emph{{Generalized Threshold
  Factorization with Full Collinear Dynamics}},
  \href{https://arxiv.org/abs/1908.00985}{{\ttfamily 1908.00985}}.

\bibitem{Bonvini:2023mfj}
M.~Bonvini and G.~Marinelli, \emph{{On the approaches to threshold resummation
  of rapidity distributions for the Drell-Yan process}},
  \href{https://doi.org/10.1140/epjc/s10052-023-12089-3}{\emph{Eur. Phys. J. C}
  {\bfseries 83} (2023) 931}
  [\href{https://arxiv.org/abs/2306.03568}{{\ttfamily 2306.03568}}].

\bibitem{Ahmed:2014cha}
T.~Ahmed, N.~Rana and V.~Ravindran, \emph{{Higgs boson production through $b
  \bar b$ annihilation at threshold in N$^3$LO QCD}},
  \href{https://doi.org/10.1007/JHEP10(2014)139}{\emph{JHEP} {\bfseries 10}
  (2014) 139} [\href{https://arxiv.org/abs/1408.0787}{{\ttfamily 1408.0787}}].

\bibitem{Das:2023bfi}
G.~Das, \emph{{$Z, W^{\pm}$ rapidity distributions at NNLL and beyond}},
  \href{https://arxiv.org/abs/2303.16578}{{\ttfamily 2303.16578}}.

\bibitem{Buckley:2014ana}
A.~Buckley, J.~Ferrando, S.~Lloyd, K.~Nordstr\"om, B.~Page, M.~R\"ufenacht
  et~al., \emph{{LHAPDF6: parton density access in the LHC precision era}},
  \href{https://doi.org/10.1140/epjc/s10052-015-3318-8}{\emph{Eur. Phys. J. C}
  {\bfseries 75} (2015) 132} [\href{https://arxiv.org/abs/1412.7420}{{\ttfamily
  1412.7420}}].

\bibitem{Kulesza:2002rh}
A.~Kulesza, G.~F. Sterman and W.~Vogelsang, \emph{{Joint resummation in
  electroweak boson production}},
  \href{https://doi.org/10.1103/PhysRevD.66.014011}{\emph{Phys. Rev. D}
  {\bfseries 66} (2002) 014011}
  [\href{https://arxiv.org/abs/hep-ph/0202251}{{\ttfamily hep-ph/0202251}}].

\bibitem{Catani:1996yz}
S.~Catani, M.~L. Mangano, P.~Nason and L.~Trentadue, \emph{{The Resummation of
  soft gluons in hadronic collisions}},
  \href{https://doi.org/10.1016/0550-3213(96)00399-9}{\emph{Nucl. Phys. B}
  {\bfseries 478} (1996) 273}
  [\href{https://arxiv.org/abs/hep-ph/9604351}{{\ttfamily hep-ph/9604351}}].

\bibitem{Boughezal:2015dva}
R.~Boughezal, C.~Focke, X.~Liu and F.~Petriello, \emph{{$W$-boson production in
  association with a jet at next-to-next-to-leading order in perturbative
  QCD}}, \href{https://doi.org/10.1103/PhysRevLett.115.062002}{\emph{Phys. Rev.
  Lett.} {\bfseries 115} (2015) 062002}
  [\href{https://arxiv.org/abs/1504.02131}{{\ttfamily 1504.02131}}].

\bibitem{Gaunt:2015pea}
J.~Gaunt, M.~Stahlhofen, F.~J. Tackmann and J.~R. Walsh, \emph{{N-jettiness
  Subtractions for NNLO QCD Calculations}},
  \href{https://doi.org/10.1007/JHEP09(2015)058}{\emph{JHEP} {\bfseries 09}
  (2015) 058} [\href{https://arxiv.org/abs/1505.04794}{{\ttfamily
  1505.04794}}].

\bibitem{Campbell:1999ah}
J.~M. Campbell and R.~K. Ellis, \emph{{An Update on vector boson pair
  production at hadron colliders}},
  \href{https://doi.org/10.1103/PhysRevD.60.113006}{\emph{Phys. Rev. D}
  {\bfseries 60} (1999) 113006}
  [\href{https://arxiv.org/abs/hep-ph/9905386}{{\ttfamily hep-ph/9905386}}].

\bibitem{Campbell:2015qma}
J.~M. Campbell, R.~K. Ellis and W.~T. Giele, \emph{{A Multi-Threaded Version of
  MCFM}}, \href{https://doi.org/10.1140/epjc/s10052-015-3461-2}{\emph{Eur.
  Phys. J. C} {\bfseries 75} (2015) 246}
  [\href{https://arxiv.org/abs/1503.06182}{{\ttfamily 1503.06182}}].

\bibitem{Boughezal:2016wmq}
R.~Boughezal, J.~M. Campbell, R.~K. Ellis, C.~Focke, W.~Giele, X.~Liu et~al.,
  \emph{{Color singlet production at NNLO in MCFM}},
  \href{https://doi.org/10.1140/epjc/s10052-016-4558-y}{\emph{Eur. Phys. J. C}
  {\bfseries 77} (2017) 7} [\href{https://arxiv.org/abs/1605.08011}{{\ttfamily
  1605.08011}}].

\bibitem{Gross:1973id}
D.~J. Gross and F.~Wilczek, \emph{{Ultraviolet Behavior of Nonabelian Gauge
  Theories}}, \href{https://doi.org/10.1103/PhysRevLett.30.1343}{\emph{Phys.
  Rev. Lett.} {\bfseries 30} (1973) 1343}.

\bibitem{Politzer:1973fx}
H.~D. Politzer, \emph{{Reliable Perturbative Results for Strong
  Interactions?}},
  \href{https://doi.org/10.1103/PhysRevLett.30.1346}{\emph{Phys. Rev. Lett.}
  {\bfseries 30} (1973) 1346}.

\bibitem{Caswell:1974gg}
W.~E. Caswell, \emph{{Asymptotic Behavior of Nonabelian Gauge Theories to Two
  Loop Order}}, \href{https://doi.org/10.1103/PhysRevLett.33.244}{\emph{Phys.
  Rev. Lett.} {\bfseries 33} (1974) 244}.

\bibitem{Jones:1974mm}
D.~R.~T. Jones, \emph{{Two Loop Diagrams in Yang-Mills Theory}},
  \href{https://doi.org/10.1016/0550-3213(74)90093-5}{\emph{Nucl. Phys. B}
  {\bfseries 75} (1974) 531}.

\bibitem{Egorian:1978zx}
E.~Egorian and O.~V. Tarasov, \emph{{Two Loop Renormalization of the {QCD} in
  an Arbitrary Gauge}}, {\emph{Teor. Mat. Fiz.} {\bfseries 41} (1979) 26}.

\bibitem{Tarasov:1980au}
O.~V. Tarasov, A.~A. Vladimirov and A.~Y. Zharkov, \emph{{The Gell-Mann-Low
  Function of QCD in the Three Loop Approximation}},
  \href{https://doi.org/10.1016/0370-2693(80)90358-5}{\emph{Phys. Lett. B}
  {\bfseries 93} (1980) 429}.

\bibitem{Larin:1993tp}
S.~A. Larin and J.~A.~M. Vermaseren, \emph{{The Three loop QCD Beta function
  and anomalous dimensions}},
  \href{https://doi.org/10.1016/0370-2693(93)91441-O}{\emph{Phys. Lett. B}
  {\bfseries 303} (1993) 334}
  [\href{https://arxiv.org/abs/hep-ph/9302208}{{\ttfamily hep-ph/9302208}}].

\bibitem{vanRitbergen:1997va}
T.~van Ritbergen, J.~A.~M. Vermaseren and S.~A. Larin, \emph{{The Four loop
  beta function in quantum chromodynamics}},
  \href{https://doi.org/10.1016/S0370-2693(97)00370-5}{\emph{Phys. Lett. B}
  {\bfseries 400} (1997) 379}
  [\href{https://arxiv.org/abs/hep-ph/9701390}{{\ttfamily hep-ph/9701390}}].

\bibitem{Czakon:2004bu}
M.~Czakon, \emph{{The Four-loop QCD beta-function and anomalous dimensions}},
  \href{https://doi.org/10.1016/j.nuclphysb.2005.01.012}{\emph{Nucl. Phys. B}
  {\bfseries 710} (2005) 485}
  [\href{https://arxiv.org/abs/hep-ph/0411261}{{\ttfamily hep-ph/0411261}}].

\bibitem{Baikov:2016tgj}
P.~A. Baikov, K.~G. Chetyrkin and J.~H. K\"uhn, \emph{{Five-Loop Running of the
  QCD coupling constant}},
  \href{https://doi.org/10.1103/PhysRevLett.118.082002}{\emph{Phys. Rev. Lett.}
  {\bfseries 118} (2017) 082002}
  [\href{https://arxiv.org/abs/1606.08659}{{\ttfamily 1606.08659}}].

\bibitem{Herzog:2017ohr}
F.~Herzog, B.~Ruijl, T.~Ueda, J.~A.~M. Vermaseren and A.~Vogt, \emph{{The
  five-loop beta function of Yang-Mills theory with fermions}},
  \href{https://doi.org/10.1007/JHEP02(2017)090}{\emph{JHEP} {\bfseries 02}
  (2017) 090} [\href{https://arxiv.org/abs/1701.01404}{{\ttfamily
  1701.01404}}].

\bibitem{Luthe:2017ttg}
T.~Luthe, A.~Maier, P.~Marquard and Y.~Schroder, \emph{{The five-loop Beta
  function for a general gauge group and anomalous dimensions beyond Feynman
  gauge}}, \href{https://doi.org/10.1007/JHEP10(2017)166}{\emph{JHEP}
  {\bfseries 10} (2017) 166}
  [\href{https://arxiv.org/abs/1709.07718}{{\ttfamily 1709.07718}}].

\bibitem{Hahn:2004fe}
T.~Hahn, \emph{{CUBA: A Library for multidimensional numerical integration}},
  \href{https://doi.org/10.1016/j.cpc.2005.01.010}{\emph{Comput. Phys. Commun.}
  {\bfseries 168} (2005) 78}
  [\href{https://arxiv.org/abs/hep-ph/0404043}{{\ttfamily hep-ph/0404043}}].

\bibitem{Hahn:2014fua}
T.~Hahn, \emph{{Concurrent Cuba}},
  \href{https://doi.org/10.1088/1742-6596/608/1/012066}{\emph{J. Phys. Conf.
  Ser.} {\bfseries 608} (2015) 012066}
  [\href{https://arxiv.org/abs/1408.6373}{{\ttfamily 1408.6373}}].

\bibitem{Das:2019bxi}
G.~Das, M.~C. Kumar and K.~Samanta, \emph{{Resummed inclusive cross-section in
  ADD model at N$^{3}$LL}},
  \href{https://doi.org/10.1007/JHEP10(2020)161}{\emph{JHEP} {\bfseries 10}
  (2020) 161} [\href{https://arxiv.org/abs/1912.13039}{{\ttfamily
  1912.13039}}].

\bibitem{Das:2020gie}
G.~Das, M.~C. Kumar and K.~Samanta, \emph{{Resummed inclusive cross-section in
  Randall-Sundrum model at NNLO+NNLL}},
  \href{https://doi.org/10.1007/JHEP07(2020)040}{\emph{JHEP} {\bfseries 07}
  (2020) 040} [\href{https://arxiv.org/abs/2004.03938}{{\ttfamily
  2004.03938}}].

\bibitem{Das:2020pzo}
G.~Das, M.~C. Kumar and K.~Samanta, \emph{{Precision QCD phenomenology of
  exotic spin-2 search at the LHC}},
  \href{https://doi.org/10.1007/JHEP04(2021)111}{\emph{JHEP} {\bfseries 04}
  (2021) 111} [\href{https://arxiv.org/abs/2011.15121}{{\ttfamily
  2011.15121}}].

\bibitem{Bizon:2018foh}
W.~Bizo\'n, X.~Chen, A.~Gehrmann-De~Ridder, T.~Gehrmann, N.~Glover, A.~Huss
  et~al., \emph{{Fiducial distributions in Higgs and Drell-Yan production at
  N$^{3}$LL+NNLO}}, \href{https://doi.org/10.1007/JHEP12(2018)132}{\emph{JHEP}
  {\bfseries 12} (2018) 132}
  [\href{https://arxiv.org/abs/1805.05916}{{\ttfamily 1805.05916}}].

\bibitem{Vermaseren:2000nd}
J.~A.~M. Vermaseren, \emph{{New features of FORM}},
  \href{https://arxiv.org/abs/math-ph/0010025}{{\ttfamily math-ph/0010025}}.

\bibitem{Ruijl:2017dtg}
B.~Ruijl, T.~Ueda and J.~Vermaseren, \emph{{FORM version 4.2}},
  \href{https://arxiv.org/abs/1707.06453}{{\ttfamily 1707.06453}}.

\bibitem{Henn:2019swt}
J.~M. Henn, G.~P. Korchemsky and B.~Mistlberger, \emph{{The full four-loop cusp
  anomalous dimension in $\mathcal{N}=4$ super Yang-Mills and QCD}},
  \href{https://doi.org/10.1007/JHEP04(2020)018}{\emph{JHEP} {\bfseries 04}
  (2020) 018} [\href{https://arxiv.org/abs/1911.10174}{{\ttfamily
  1911.10174}}].

\bibitem{Huber:2019fxe}
T.~Huber, A.~von Manteuffel, E.~Panzer, R.~M. Schabinger and G.~Yang,
  \emph{{The four-loop cusp anomalous dimension from the $N=4$ Sudakov form
  factor}}, \href{https://doi.org/10.1016/j.physletb.2020.135543}{\emph{Phys.
  Lett. B} {\bfseries 807} (2020) 135543}
  [\href{https://arxiv.org/abs/1912.13459}{{\ttfamily 1912.13459}}].

\bibitem{vonManteuffel:2020vjv}
A.~von Manteuffel, E.~Panzer and R.~M. Schabinger, \emph{{Cusp and collinear
  anomalous dimensions in four-loop QCD from form factors}},
  \href{https://doi.org/10.1103/PhysRevLett.124.162001}{\emph{Phys. Rev. Lett.}
  {\bfseries 124} (2020) 162001}
  [\href{https://arxiv.org/abs/2002.04617}{{\ttfamily 2002.04617}}].

\bibitem{Tarasov:1982plg}
O.~V. Tarasov, \emph{{Anomalous dimensions of quark masses in the three-loop
  approximation}}, \href{https://doi.org/10.1134/S1547477120020223}{\emph{Phys.
  Part. Nucl. Lett.} {\bfseries 17} (2020) 109}
  [\href{https://arxiv.org/abs/1910.12231}{{\ttfamily 1910.12231}}].

\bibitem{Larin:1993tq}
S.~A. Larin, \emph{{The Renormalization of the axial anomaly in dimensional
  regularization}},
  \href{https://doi.org/10.1016/0370-2693(93)90053-K}{\emph{Phys. Lett. B}
  {\bfseries 303} (1993) 113}
  [\href{https://arxiv.org/abs/hep-ph/9302240}{{\ttfamily hep-ph/9302240}}].

\bibitem{Alekseev:1994xp}
E.~N. Alekseev, V.~A. Matveev, K.~S. Nirov and V.~A. Rubakov, eds.,
  \emph{{Particles and cosmology. Proceedings, International School, Baksan
  Valley, Russia, April 22-27, 1993}}, 1994.

\bibitem{Chetyrkin:1997dh}
K.~G. Chetyrkin, \emph{{Quark mass anomalous dimension to ${\cal O}
  (\alpha_S^4)$}},
  \href{https://doi.org/10.1016/S0370-2693(97)00535-2}{\emph{Phys. Lett. B}
  {\bfseries 404} (1997) 161}
  [\href{https://arxiv.org/abs/hep-ph/9703278}{{\ttfamily hep-ph/9703278}}].

\end{thebibliography}\endgroup
\end{document}